\documentclass[prd,preprintnumbers,nofootinbib,twocolumn,longbibliography]{revtex4-1}
\usepackage{graphicx}
\usepackage{amsmath}
\usepackage{amssymb}
\usepackage{braket}
\usepackage{slashed}
\usepackage{color}
\usepackage[bookmarksopen,bookmarksnumbered]{hyperref}
\usepackage{fix-revtex-bib}
\allowdisplaybreaks

\newcommand\f[2]{\frac{#1}{#2}} 

\def\xleft{\mathopen{}\left}

\DeclareRobustCommand*\diff[2][]{%
   \mathop{
     \mathrm{d}^{#1}
     \mskip-0.2\thinmuskip
    #2}\nolimits
}

\newcommand\3[1]{\boldsymbol{#1}}

\newcommand{\T}[1]{\boldsymbol{#1}_{\text{T}}}
\newcommand{\Tsc}[1]{#1_{\text{T}}}

\newcommand\BIBvol[1]{\textbf{#1}}

\newcommand{\bmax}{b_{\rm max}}
\newcommand\bstar{\3{b}_*}
\newcommand\bstarsc{b_*}
\newcommand\mubstar{\mu_{\bstarsc}}
\newcommand\muQ{\mu_Q}
\newcommand{\gammae}{\ensuremath{\gamma_{\rm E}}}

\def\??#1{\textbf{\textcolor{red}{---??#1---}}}
\newcommand{\arrowcom}[1]{\textcolor{blue}{ \textbf{$\Longrightarrow$ #1}} \\}
\def\!!#1{\mbox{}\\\arrowcom{#1}}

\DeclareRobustCommand{\MSbar}{\ensuremath{ \overline{\rm MS} }}

\begin{document}


\title{Connecting Different TMD Factorization Formalisms in QCD}

\preprint{JLAB-THY-17-2470}    
  
\author{John Collins}
\email{jcc8@psu.edu}
\affiliation{%
  Department of Physics, Penn State University, University Park PA 16802, USA}
\author{Ted C. Rogers}
\email{trogers@odu.edu}
\affiliation{Theory Center, Jefferson Lab, 12000 Jefferson Avenue, Newport News, 
VA 23606, USA}
\affiliation{Department of Physics, Old Dominion University, Norfolk, VA 23529, 
USA}  
\date{1 June 2017}

\begin{abstract}
  In the original Collins-Soper-Sterman (CSS) presentation of the results of
  transverse-momentum-dependent (TMD) factorization for the Drell-Yan
  process, results for perturbative coefficients can be obtained from
  calculations for collinear factorization.  Here we show how to use
  these results, plus known results for the quark form factor, to
  obtain coefficients for TMD factorization in more recent
  formulations, e.g., that due to Collins, and apply them to known
  results at order $\alpha_s^2$ and $\alpha_s^3$.  We also show that the
  ``non-perturbative'' functions as obtained from fits to data are
  equal in the two schemes. We compile the higher-order perturbative inputs needed for the updated CSS scheme by appealing to results obtained in a 
  variety of different formalisms. In addition, we derive the connection between both versions of the CSS formalism and several formalisms based 
  in soft-collinear effective theory (SCET). Our work uses some important new results for
  factorization for the quark form factor, which we derive.
\end{abstract}

\maketitle

\section{Introduction}
\label{sec:intro}

In the application of transverse-momentum-dependent (TMD)
factorization to the Drell-Yan and other processes,
many standard fits to
data, like those of Refs.\ \cite{Landry:2002ix, Konychev:2005iy}, use
a presentation of a TMD factorization formula due to Collins, Soper,
and Sterman (CSS1) \cite{Collins:1984kg}.  In this method, the cross
section is written as a Fourier transform over a transverse-position
variable $\T{b}$.  The $\T{b}$ dependence is separated into a part
estimated by perturbative methods, and a correction factor involving
certain functions $g_K(\Tsc{b})$, $g_{j/H}(x,\Tsc{b})$ that allow for
a parametrization of the important non-perturbative dependence at
large $\Tsc{b}$.  The perturbative part is restricted to use $\Tsc{b}$
less than a cut-off $\bmax$.  Results of fits are presented as
parametrizations for the ``non-perturbative'' functions\footnote{The
  characterization of these functions as non-perturbative is somewhat
  misleading.  While the intent of their definition is to include the
  important non-perturbative properties of TMD functions at large
  $\Tsc{b}$, they can also include perturbatively calculable
  contributions if $\bmax$ is chosen conservatively small. }
$g_K(\Tsc{b})$, $g_{j/H}(x,\Tsc{b}{})$ etc. (Let us call these 
collectively the ``$g$-functions.'')

Two issues now arise.  The first is that an improved version of TMD
factorization has been derived in Ref.\
  \cite{Collins:2011qcdbook}, and that some closely related formalisms
  have been developed within the framework of soft-collinear effective
  theory (SCET)\footnote{We will comment on some of the
      relations of the SCET-based formalisms to CSS2 in Sec.\
      \ref{sec:CSS2} and in App.\ \ref{app:corr.Li.et.al}.  In
      particular, the TMD functions defined by
      \citet{GarciaEchevarria:2011rb} agree with those of CSS2, as
      does the way in which they appear in the TMD factorization
      formula.  In Sec.\ \ref{sec:mismatch}, we will summarizes the
      relevant differences between CSS1 and the newer methods.
}.  Let us refer to the version in Ref.\
  \cite{Collins:2011qcdbook} as CSS2.  

The second issue is that
the fitted functions, the $g$-functions, are not intrinsically
interpretable in terms of TMD parton densities, but only in
conjunction with the 
cut-off-dependent perturbative part of the
factorization formula. This raises questions about the validity of using 
$g$-functions extracted using one perturbative formalism for calculations and 
phenomenology in another formalism. \citet{Aybat:2011zv} already organized
the TMD functions in accordance with the new definitions,
and used existing previously existing phenomenology to construct TMD
parametrizations of parton densities in terms of $g$-functions.
However, until now it has not been firmly established that the $g$-functions 
extracted using the older CSS1 formalism actually apply directly to the TMD 
functions defined in CSS2.

In this article we therefore do the following: We show how to relate
the two versions of the CSS-style formalism, so that results of fits
obtained using the original CSS factorization formula can be applied
in the new formalism.  We also derive explicit transformations to
implement the scheme change between the two formalisms.  Key
quantities in both formalisms are TMD parton densities and the CSS
evolution kernel $\tilde{K}(\Tsc{b})$, which are defined in terms of
certain QCD matrix elements.  The present article's advances include obtaining 
the full relations between the old and new schemes, showing completely how fits 
made using the old scheme can be applied to give TMD parton densities in the 
new scheme. We show that the $g$-functions
in the two schemes are equal. We give formulae for the TMD functions 
with the new definitions in terms of the fitted functions obtained using 
the original CSS formalism.  The resulting TMD functions have an invariant 
significance, independently of the details of the specific implementation and of
values of arbitrary perturbative cutoffs like the renormalization 
scale and the $\bmax$ cut off.

We compute various functions needed in the formalism, on the basis of
existing calculations of the quark form factor by \citet{Moch:2005id},
and of hard scattering in collinear factorization by
\citet{Catani:2000vq}. These results are: (a) The coefficients
relating TMD and collinear parton densities to order $a_s^2$;  (b) The
TMD hard scattering coefficient for Drell-Yan to order $a_s^2$;  (c)
The anomalous dimensions to order $a_s^3$; (d) The CSS2 evolution
kernel $\tilde{K}$ to order $a_s^2$.  We give full details of the
non-trivial methods by which the coefficients are obtained from the
previous results.  In particular we find that we need some apparently
new technical results concerning the collinear factors used
for factorization for the quark form factor.  We verify that our
results agree with calculations of corresponding quantities by very
different methods by \citet{Gehrmann:2012ze, Gehrmann:2014yya} and by
\citet{Echevarria:2016scs}.  Those calculations start from the
operator definitions of the TMD functions, and so the agreement with
our calculations provides a non-trivial test of the correctness of
the TMD factorization methods.  We point out that the order $a_s^3$
value for the hard scattering is available from results by
\citet{Gehrmann:2010ue}, and that a calculation by \citet{Li:2016ctv}
gives the value of $\tilde{K}$ to order $a_s^3$.  That the result of
Ref.\ \cite{Li:2016ctv} in fact gives exactly the perturbative
expansion of $\tilde{K}$ is not immediately apparent from their paper,
so we give a derivation of the correspondence in App.\
\ref{app:corr.Li.et.al}, where we also show how to map their
factorization and TMD parton densities onto those given by CSS2 and by
\citet{GarciaEchevarria:2011rb}.

\section{The formalisms}
\label{sec:CSS}

\subsection{Notation and conventions}

To match the conventions of \citet{Moch:2005id}, we use
\begin{equation}
\label{eq:as.def}
  a_s = \frac{\alpha_s}{4\pi} = \frac{g_s^2}{ 16 \pi^2 }
\end{equation}
as the expansion parameter.

\subsection{Original CSS formalism}
\label{sec:CSS1}

The original CSS formula \cite[(3.17) and (5.8)]{Collins:1984kg}, as
used in the fits in \cite{Landry:2002ix, Konychev:2005iy}, was
obtained starting from a TMD factorization formula, using the
  specific definitions of
TMD parton densities that had been given by Collins and Soper
(CS) 
\cite{Collins:1981uw}.  Earlier, CS \cite{Collins:1981uk,
  Collins:1981va} had obtained TMD factorization for dihadron
production in $e^+e^-$ annihilation.  The natural extension to the
Drell-Yan process was stated by CSS in \cite{Collins:1984kg}; CSS
argued that the then-recent work on the cancellation of the Glauber
region was sufficient to allow the extension of the proof of TMD
factorization to Drell-Yan.

Associated with factorization are evolution equations for the TMD
functions and a kind of operator-product expansion (OPE) for the TMD
parton densities at small $\Tsc{b}$. CSS solved these equations
with neglect of power-suppressed terms,
segregated non-perturbative contributions at large $\Tsc{b}$, and then
redefined various functions.  The result was
of the form
\begin{widetext}
\begin{align}
  \label{eq:CSS.A.B}
  \frac{\diff{\sigma}}{\diff{Q^2}\diff{y}\diff{\Tsc{q}^2}}
  ={}&
   \frac{4\pi^2\alpha^2}{9Q^2s} \sum_{j,j_A,j_B} e_j^2
   \int \frac{ \diff[2]{\T{b}} }{ (2\pi)^2 } e^{i\T{q}\cdot\T{b}}
\nonumber\\& \times
  \int_{x_A}^1 \frac{ \diff{\xi_A} }{ \xi_A }
       f_{j_A/A}(\xi_A;\mubstar) 
  ~ \tilde{C}^{\text{CSS1, DY}}_{j/j_A}\xleft( \frac{x_A}{\xi_A},\bstarsc;
  \mubstar^2, \mubstar, C_2, a_s(\mubstar) \right)
\nonumber\\& \times
  \int_{x_B}^1 \frac{ \diff{\xi_B} }{ \xi_B }
       f_{j_B/B}(\xi_B;\mubstar) 
  ~ \tilde{C}^{\text{CSS1, DY}}_{\bar{\jmath}/j_B}\xleft(
             \frac{x_B}{\xi_B},\bstarsc; \mubstar^2, \mubstar, C_2, a_s(\mubstar) \right)
\nonumber\\& \times
  \exp\left\{
    - \int_{\mubstar^2}^{\muQ^2}
      \frac{ \diff{ {\mu'}^2 } }{ {\mu'}^2 }
      \left[
             A_{\rm CSS1}(a_s(\mu'); C_1 ) \ln\left( \frac{\muQ^2}{{\mu'}^2} \right)
             + B_{\text{CSS1, DY}}(a_s(\mu');C_1,C_2 )
      \right]
  \right\}
\nonumber\\& \times  
  \exp\xleft[ - g^{\rm CSS1}_{j/A}(x_A,\Tsc{b};\bmax)
              - g^{\rm CSS1}_{\bar{\jmath}/B}(x_B,\Tsc{b};\bmax)
              - g^{\rm CSS1}_K(\Tsc{b};\bmax) \ln(Q^2/Q_0^2)
      \right]
\nonumber\\
  & + \mbox{suppressed corrections}.
\end{align}
\end{widetext}
Here we work with the inclusive Drell-Yan process $A+B\to l^+l^- + X$,
with restriction to production of the lepton pair through a virtual
photon.  The 4-momentum of the lepton pair is $q^\mu$, and its
invariant mass, rapidity and transverse momentum are $Q$, $y$ and
$\T{q}$.  The total center of mass energy is $\sqrt{s}$, we define
$x_A = Qe^y/\sqrt{s}$ and $x_B = Qe^{-y}/\sqrt{s}$,
we define $e_j$ to be the charge of quark $j$ (in units of the
elementary charge unit $e$), and $\alpha$ is the usual fine-structure
constant.  Auxiliary quantities are defined by
\begin{align}
\label{eq:bstar}
  \bstar & = \frac{ \T{b} }{ \sqrt{ 1 + \Tsc{b}^2/\bmax^2} },
\\
  \label{eq:mub.defn}
  \mubstar &= C_1 / \bstarsc,
\\
  \label{eq:muQ.defn}
   \muQ &= C_2 Q,
\end{align}
with $C_1$ and $C_2$ being constants that can be adjusted to try to
optimize the accuracy of perturbative calculations; if all quantities
were computed exactly, the results of predictions would be independent
of $C_1$ and $C_2$.  The quantities $f_{j/H}$ are ordinary collinear
parton densities (in the $\MSbar$ scheme, normally).
Those quantities
that are specific to the particular definitions given by CSS are
indicated with the label ``CSS1''.  The functions $A_{\rm CSS1}$,
$B_{\text{CSS1, DY}}$, and $\tilde{C}^{\text{CSS1, DY}}$ are
perturbatively calculable.\footnote{There are two apparently
  redundant arguments for $\tilde{C}^{\text{CSS1, DY}}$ that both
  involve $\mubstar$. These correspond to the two kinds of scale
  arguments, $\zeta$ and $\mu$, for TMD functions in the CSS formalism, but
  set to appropriate values for perturbative calculations after use of
  the evolution equations.}
Corrections to the formula, as noted on the last line, are power
suppressed when $Q$ is large and $\Tsc{q} \ll Q$.  We will ignore
various polarization-related effects that have considerable current
interest (e.g., Ref.\ \cite{Adolph:2014zba, Aubert:2015hha,
  Ablikim:2015sma})
but that do not directly intersect with the issues 
that are the main concern of this article; our results can be straightforwardly extended to 
the case of polarization dependent observables.

The derivation of (\ref{eq:CSS.A.B}) from the underlying TMD
factorization formula used a certain set of redefinitions
\cite{Collins:1981va, Collins:1984kg} of various parts of the
factorization formula. An important motivation was to express the
cross section in terms of quantities that can be related to
experimental data.  For example, in the initial CSS1 factorization
formula there is a soft factor.  This has non-perturbative
contributions but always appears multiplying a pair of TMD parton
densities or fragmentation functions.  Thus, the non-perturbative part
of the soft factor cannot be separately and unambiguously 
deduced from data, even in principle. A properly defined soft factor is universal 
between reactions \cite[Ch.\ 13]{Collins:2011qcdbook}.  So absorbing a square
root of the soft factor into each parton density and fragmentation
function is sensible.

Having separate and explicitly defined TMD pdf definitions also opens the 
possibility to study such objects non-perturbatively, e.g., with lattice 
QCD~\cite{Engelhardt:2016dfw}.

But CSS1 also absorbed a square root of a hard factor into the parton
densities and fragmentation functions.  This is much less desirable,
since these functions then become process dependent---see
\cite{Catani:2000vq} and \cite[p.\ 455]{Collins:1981va}.
The hard scattering is always perturbatively 
calculable (and hence predictable), so the CSS1 procedure obscures the
predictable differences between processes. 

The new method of CSS2, reviewed in the next section, is better from this 
point of view.

\subsection{New TMD formalism}
\label{sec:CSS2}

\citet[Chs.\ 10 \& 13]{Collins:2011qcdbook} provided an updated
TMD factorization.  Much more complete derivations were provided.
Relative to CSS1, the most notable change is a modified definition of
the TMD parton densities and fragmentation functions, in terms of
explicit gauge-invariant operator matrix elements.  The new definitions have as a
consequence that the TMD factorization formula no longer contains an
explicit soft factor.  Furthermore, the definitions were arranged so
that the evolution equations are exact in their form, instead of
having power-suppressed corrections; this makes the relation between
the results of fits and the actual TMD parton densities much more
transparent.

The TMD functions, with their new definitions, are demonstrably
process independent, up to possible sign changes associated with T-odd 
functions. In the factorization formula, only the perturbatively calculable hard
scattering contains process dependence.  The method also avoids the
divergences that were found by \citet[App.\
A]{Bacchetta:2008xw} when the original CS definition of TMD densities
is taken literally.

Despite these changes, the new method should be considered a scheme
change relative the original CS/CSS definitions, as we will see in
later sections.  

A summary of the new method can be found in \cite{Collins:2014jpa},
together with a set of different forms of solution.  Here, we will
present only those results needed for our purposes, but adapted to the
cross section given in Eq.\ (\ref{eq:CSS.A.B}).

Within the framework of SCET, closely related TMD factorization
results have been given by \citet{Becher:2010tm} and
by \citet{GarciaEchevarria:2011rb}.  The results
of Echevarr\'\i a et al.\ are equivalent \cite{Collins:2012uy} to
those presented here, with the TMD functions being the same (up to
possible elementary changes in the scheme used for UV
renormalization); their formula defining the TMD densities is simpler
than that of Ref.\ \cite{Collins:2011qcdbook}.  Becher and Neubert did not define 
separately finite TMD functions.  But they did define the product of two such
functions, as used in factorization formulae, and the product agrees
with the product of the TMD functions used here and by Echevarr\'\i a
et al.  (Details of this can be extracted from a comparison of the
relevant formulae in \cite{Becher:2010tm, GarciaEchevarria:2011rb,
  Collins:2012uy}.)
There is also the formulation of TMD factorization given by
  \citet{Li:2016axz}, which looks rather different.  We will show in
  App.\ \ref{app:corr.Li.et.al} how it can be mapped, non-trivially,
  onto the CSS2 formalism; the result will enable use in CSS2 of the
  order $a_s^3$ calculations of the evolution of the soft factor that
  were given by \citet{Li:2016ctv}.

The TMD factorization formula is
\begin{widetext}
\begin{align}
\label{eq:fact.CSS2}
  \frac{\diff{\sigma}}{\diff{Q^2}\diff{y}\diff{\Tsc{q}^2}}
  ={}&
    \frac{4\pi^2\alpha^2}{9Q^2s}
    \sum_j H^{\rm DY}_{j\bar{\jmath}}(Q, \muQ, a_s(\muQ))
    \int \frac{ \diff[2]{\T{b}} }{ (2\pi)^2 }
    ~ e^{i\T{q}\cdot \T{b} }
    ~ \tilde{f}_{j/A}(x_A,\Tsc{b};Q^2,\muQ) 
    ~ \tilde{f}_{\bar{\jmath}/B}(x_B,\Tsc{b};Q^2,\muQ)
\nonumber\\
  & + \mbox{suppressed corrections},
\end{align}
where the hard scattering factor $H^{\rm DY}_{j\bar{\jmath}}$ is normalized so
that its lowest order term is $e_j^2$.
The scale argument of $H$ is set to $\muQ$ to avoid large
logarithms.  The last two arguments of the parton densities,
$f_{j/H}(x,\Tsc{b};Q^2,\muQ)$, are normally written as $\zeta$ and
$\mu$, and these arguments refer to effective cutoffs on rapidity and
transverse momentum as implemented by the definitions in
\cite{Collins:2011qcdbook}.

Predictions are obtained with the aid of evolution equations and the
small-$\Tsc{b}$ OPE of the TMD parton densities:
\begin{align}
\label{eq:CSS.evol}
  \frac{ \partial \ln \tilde{f}_{f/H}(x,\Tsc{b}; \zeta; \mu) }
       { \partial \ln \sqrt{\zeta} }
  = {}&
  \tilde{K}(\Tsc{b};\mu).
\\
\label{eq:RG.K}
  \frac{ \diff{\tilde{K}(\Tsc{b};\mu)} }{ \diff{\ln \mu } }
  = {}& -\gamma_K\xleft(a_s(\mu)\right), 
\\
\label{eq:RG.TMD.pdf}
  \frac{ \diff{ \ln \tilde{f}_{j/H}(x,\Tsc{b};\zeta;\mu) }}
       { \diff{\ln \mu} }
   = {}& \gamma_j( a_s(\mu))
      - \frac12 \gamma_K(a_s(\mu)) \ln \frac{ \zeta }{ \mu^2 },
\\
\label{eq:TMD.OPE}
  \tilde{f}_{j/H}(x,\Tsc{b};\zeta;\mu) 
  = {}& \sum_k \int_{x-}^{1+} \frac{ \diff{\xi} }{ \xi }
       \,\tilde{C}^{\rm PDF}_{j/k}\xleft( x/\xi,\Tsc{b};\zeta,\mu,a_s(\mu) \right)
        f_{k/H}(\xi;\mu)
~+~ O\xleft[(m\Tsc{b})^p \right].
\end{align}
(For an explanation of the notations $x-$ and $1+$ for 
the integration limits, see \cite[pp.\ 248 \& 249]{Collins:2011qcdbook}.)
A solution that corresponds to Eq.\ (\ref{eq:CSS.A.B}) is
\begin{align}
\label{eq:soln.2}
  \frac{\diff{\sigma}}{\diff{Q^2}\diff{y}\diff{\Tsc{q}^2}}
  ={}&
    \frac{4\pi^2\alpha^2}{9Q^2s}
    \sum_{j,j_A,j_B} H^{\rm DY}_{j\bar{\jmath}}(Q, \muQ, a_s(\muQ))
        \int \frac{ \diff[2]{\T{b}} }{ (2\pi)^2 }  e^{i\T{q}\cdot \T{b} }
\nonumber\\& ~~ \times
  e^{-g_{j/A}(x_A,\Tsc{b};\bmax) }
  \int_{x_A}^1 \frac{ \diff{\xi_A} }{ \xi_A }
       f_{j_A/A}(\xi_A;\mubstar) 
  ~ \tilde{C}^{\rm PDF}_{j/j_A}\xleft( \frac{x_A}{\xi_A},\bstarsc;
  \mubstar^2, \mubstar, a_s(\mubstar) \right)
\nonumber\\& ~~ \times
  e^{ -g_{\bar{\jmath}/B}(x_B,\Tsc{b};\bmax)}
  \int_{x_B}^1 \frac{ \diff{\xi_B} }{ \xi_B }
       f_{j_B/B}(\xi_B;\mubstar) 
  ~ \tilde{C}^{\rm PDF}_{\bar{\jmath}/j_B}\xleft( \frac{x_B}{\xi_B},\bstarsc; \mubstar^2, \mubstar, a_s(\mubstar) \right)
\nonumber\\& ~~ \times  
  \exp\xleft\{ 
       - g_K(\Tsc{b};\bmax) \ln \frac{ Q^2 }{ Q_0^2 } 
       + \tilde{K}(\bstarsc;\mubstar) \ln \frac{ Q^2 }{ \mubstar^2 }
       + \int_{\mubstar}^{\muQ}  \frac{ \diff{\mu'} }{ \mu' }
          \left[ 2 \gamma_j(a_s(\mu')) 
                 - \ln\frac{Q^2}{ (\mu')^2 } \gamma_K(a_s(\mu'))
          \right]
  \right\}
\nonumber\\& 
  + \mbox{suppressed corrections}.
\end{align}
\end{widetext}
Analogous equations apply to fragmentation functions in processes like
semi-inclusive deeply inelastic scattering (SIDIS) and $e^+ e^-$ annihilation, with 
the same $\tilde{K}$, $\gamma_j$, and $\gamma_K$ functions. (Equality of
$\tilde{K}$ and $\gamma_K$ between the processes was proved in Ref.\
\cite{Collins:2011qcdbook}; equality of $\gamma_j$ will be proved
in our Sec.~\ref{sec:PDFvFF}.)
Note the ${\rm DY}$ label on the hard part, $H_{j\bar{\jmath}}^{\rm DY}(Q, \muQ, a_s(\muQ))$, to 
indicate that this hard part is specific to the Drell-Yan scattering
process.
We have used the notation $\tilde{C}^{\rm PDF}$ to indicate that
the corresponding coefficients will be different for fragmentation
functions.

\subsection{The mismatches between CSS1 and the new methods}
\label{sec:mismatch}

In all the methods, the primary idea is to extract the leading power
behavior in an expansion where masses and $\Tsc{q}{}$ are small relative
to $Q$.  By far the simplest form of the results for factorization is
when the leading-power expansion is used strictly; terms of
non-leading power tend to be more complicated.  A problem is that when
a strict leading power expansion is done, one obtains individual terms
that have UV and rapidity divergences not present in the original
amplitudes.  So at intermediate stages of derivations and
calculations, cutoffs (or regulators) are applied to the divergences.
All the methods are in agreement to deal with UV divergences by
renormalization, after which the UV cutoff can be removed.  The
differences between the methods concern the treatment of rapidity
divergences.

The rapidity divergences are associated with the light-like Wilson
lines that arise when the operators in the factors are defined in the
natural gauge-invariant way that arises from the leading-power
expansion, or some equivalent property.  

In CSS1, collinear factors are defined with the use of a
non-light-like axial gauge, or equivalently with non-lightlike Wilson
lines.  For example, in the case of the quark form factor, the
collinear factor would be defined by the matrix element in Eq.\
\eqref{eq:Cj.def} below, but without the limit $y_2\to-\infty$ (and the $S$
factors are moved elsewhere).  Effectively some non-leading powers are
retained.  Correspondingly, the evolution equations have power
corrections; these were not analyzed by CSS, and instead the
corrections are dropped in a solution such as Eq.\ \eqref{eq:CSS.A.B}.
Thus there is a mismatch between the actual TMD pdfs defined in Ref.\
\cite{Collins:1981uw} and those that correspond to Eq.\
\eqref{eq:CSS.A.B}, although the differences are power suppressed.

Furthermore, in CSS1 the TMD factors were then redefined to remove the
hard factor and soft factor that would otherwise be present; this
produces the process dependence of the TMD parton densities and
fragmentation functions that was mentioned earlier.

In CSS2 and the SCET methods we have quoted, a strict leading power
expansion is used.  Although cutoffs on rapidity divergences are used
at intermediate stages, these are removed at the end.  Thus the basic
collinear and soft factors have the lightlike Wilson lines that
naturally arise from a gauge-invariant implementation of the leading
power expansion.  There are then applied certain kinds of
reorganization of the factors and/or a
generalized renormalization of the rapidity divergences. 
These both avoid
double counting of the contributions of different regions and
ensure that individual TMD functions used in factorization are
finite.  

In CSS2 itself, there remain non-lightlike Wilson lines, as in Eq.\
\eqref{eq:Cj.def} below.  But these are always in matrix elements of a
basic soft factor where the other Wilson line is lightlike.  The
dependence of a collinear factor on the direction of this Wilson line
gives the $\zeta$ dependence of the TMD functions.  In contrast in the
SCET methods, especially that of \citet{GarciaEchevarria:2011rb}, the
Wilson lines are always lightlike, and another regulator is used.  The
role of the direction of CSS2's non-lightlike Wilson line is now
played by a choice of coordination between the regulator of oppositely
directed Wilson lines; it gives rise to the same $\zeta$ dependence
\cite{Collins:2012uy}.  The final factorization formula and the TMD
functions are defined in the limit that the regulators are removed.
The factorization formula then has exactly the leading power, and the
evolution equations are homogeneous without any power-suppressed
corrections.

\section{New v.\ old CSS}
\label{sec:new.old}

In this section, we show how to relate the TMD factorization
formula of CSS2 to that of CSS1.
The results are closely related to formulae given by CSS
\cite{Collins:1984kg} used in transforming their initial TMD factorization
to the form of Eq.\ (\ref{eq:CSS.A.B}).  Here we will derive the
relationship using a comparison of Eqs.\ (\ref{eq:CSS.A.B})
and (\ref{eq:soln.2}) as the starting point.

\subsection{Drell-Yan}

Both of Eqs.\ (\ref{eq:CSS.A.B}) and (\ref{eq:soln.2}) give the same
cross section.  However, they also agree for each separate term for a 
given flavor $j\bar{\jmath}$ for the annihilating quark-antiquark pair.
This is because the manipulations to get the different factorized
forms start from exactly the same graphs, and these may therefore be
restricted to those with any given quark flavor.  Once this is done,
the Fourier transform can be removed, and separate equality for each
value of $\Tsc{b}$ is obtained.  Furthermore, at least as regards what
can be seen in Feynman graphs, the two forms include exactly the
leading power. The derivations of CSS1 and CSS2 drop the same 
subleading powers to get factorization, so this equality is exact, rather than being 
merely modulo power-suppressed corrections.  Hence we have
\begin{widetext}
\begin{align}
\label{eq:CSS1.v.CSS2}
  & e_j^2 \sum_{j_A} \int_{x_A}^1 \frac{ \diff{\xi_A} }{ \xi_A }
       f_{j_A/A}(\xi_A;\mubstar) 
  ~ \tilde{C}^{\text{CSS1, DY}}_{j/j_A}\xleft( \frac{x_A}{\xi_A},\bstarsc;
                  \mubstar^2, \mubstar, C_2, a_s(\mubstar) \right)
\nonumber\\& \times
   \sum_{j_B} \int_{x_B}^1 \frac{ \diff{\xi_B} }{ \xi_B }
       f_{j_B/B}(\xi_B;\mubstar) 
  ~ \tilde{C}^\text{CSS1, DY}_{\bar{\jmath}/j_B}\xleft(
      \frac{x_B}{\xi_B},\bstarsc; \mubstar^2, \mubstar, C_2, a_s(\mubstar) \right)
\nonumber\\& \times
  \exp\left\{
    - \int_{\mubstar^2}^{\muQ^2}
      \frac{ \diff{ {\mu'}^2 } }{ {\mu'}^2 }
      \left[
             A_{\rm CSS1}(a_s(\mu'); C_1 ) \ln\left( \frac{\muQ^2}{{\mu'}^2} \right)
             + B_{\text{CSS1, DY}}(a_s(\mu');C_1,C_2 )
      \right]
  \right\}
\nonumber\\& \times  
  \exp\xleft[ - g^{\rm CSS1}_{j/A}(x_A,\Tsc{b};\bmax)
              - g^{\rm CSS1}_{\bar{\jmath}/B}(x_B,\Tsc{b};\bmax)
              - g^{\rm CSS1}_K(\Tsc{b};\bmax) \ln(Q^2/Q_0^2)
      \right]
\nonumber\displaybreak[0]\\
  ={}&
  H^{\rm DY}_{j\bar{\jmath}}(Q, \muQ, a_s(\muQ))
\nonumber\\& \times
  \sum_{j_A} \int_{x_A}^1 \frac{ \diff{\xi_A} }{ \xi_A }
       f_{j_A/A}(\xi_A;\mubstar) 
  ~ \tilde{C}^{\rm PDF}_{j/j_A}\xleft( \frac{x_A}{\xi_A},\bstarsc;
  \mubstar^2, \mubstar, a_s(\mubstar) \right)
\nonumber\\& \times
  \sum_{j_B} \int_{x_B}^1 \frac{ \diff{\xi_B} }{ \xi_B }
       f_{j_B/B}(\xi_B;\mubstar) 
  ~ \tilde{C}^{\rm PDF}_{\bar{\jmath}/j_B}\xleft( \frac{x_B}{\xi_B},\bstarsc; \mubstar^2, \mubstar, a_s(\mubstar) \right)
\nonumber\\& \times
  \exp\xleft\{ 
         \tilde{K}(\bstarsc;\mubstar) \ln \frac{ Q^2 }{ \mubstar^2 }
       + \int_{\mubstar}^{\muQ}  \frac{ \diff{\mu'} }{ \mu' }
          \left[ 2 \gamma_j(a_s(\mu')) 
                 - \ln\frac{Q^2}{ (\mu')^2 } \gamma_K(a_s(\mu'))
          \right]
  \right\}
\nonumber\\& \times  
  \exp\xleft[ - g_{j/A}(x_A,\Tsc{b};\bmax)
              - g_{\bar{\jmath}/B}(x_B,\Tsc{b};\bmax)
              - g_K(\Tsc{b};\bmax) \ln(Q^2/Q_0^2)
      \right].
\end{align}
Although there are clear structural similarities, the structures do
not exactly correspond on the two sides of this equation. 
Note that the CSS1 coefficients used here are specific to
parton densities and the Drell-Yan process.

First, we differentiate both sides with respect to all the dependence
on $\ln Q^2$.  This gives
\begin{multline}
\label{eq:CSS1.v.CSS2.Q}
    - g^{\rm CSS1}_K(\Tsc{b};\bmax) 
    - B_{\text{CSS1, DY}}(a_s(\muQ);C_1,C_2 )
    - \int_{\mubstar^2}^{\muQ^2}
      \frac{ \diff{ {\mu'}^2 } }{ {\mu'}^2 }
             A_{\rm CSS1}(a_s(\mu'); C_1 )
\\
  ={}
    - g_K(\Tsc{b};\bmax)
    + \tilde{K}(\bstarsc;\mubstar)
    + \frac{ \diff{ \ln H^{\rm DY}_{j\bar{\jmath}}(Q, \muQ, a_s(\muQ)) } }
           { \diff{\ln Q^2} }
    +  \gamma_j(a_s(\muQ)) 
    -  \ln\frac{Q}{\muQ} \gamma_K(a_s(\muQ))
    - \int_{\mubstar}^{\muQ}  \frac{ \diff{\mu'} }{ \mu' }
             \gamma_K(a_s(\mu')) \, .
\end{multline}
Then differentiating with respect to $\ln \Tsc{b}^2$ gives
\begin{align}
\label{eq:CSS1.v.CSS2.Q.b}
    \frac{ \diff{g^{\rm CSS1}_K(\Tsc{b};\bmax) } }
           { \diff{\ln\Tsc{b}^2} }
    + \frac{\bstarsc^2}{\Tsc{b}^2} 
      A_{\rm CSS1}(a_s(\mubstar); C_1 )
  ={}&
    \frac{ \diff{ g_K(\Tsc{b};\bmax) } }
           { \diff{\ln\Tsc{b}^2} }
    - \frac{\bstarsc^2}{\Tsc{b}^2} 
      \left[
         \frac{ \diff{ \tilde{K}(\bstarsc;\mubstar) } }
              { \diff{\ln\bstarsc^2} }
         - \frac{1}{2} \gamma_K(a_s(\mubstar))
      \right]
\nonumber\\
   ={}&
    \frac{ \diff{ g_K(\Tsc{b};\bmax) } }
           { \diff{\ln\Tsc{b}^2} }
    - \frac{\bstarsc^2}{\Tsc{b}^2}
      \left. 
         \frac{ \partial{ \tilde{K}(\bstarsc;\mu) } }
           { \partial{\ln\bstarsc^2} }
      \right|_{\mu \mapsto \mubstar} ,
\end{align}
where we used Eq.\ (\ref{eq:RG.K}) and
\begin{equation}
  \frac{ \diff{\ln\bstarsc^2} }{ \diff{\ln\Tsc{b}^2} }
  = \frac{\bstarsc^2}{\Tsc{b}^2} .
\end{equation}
Now each of $g_K$ and $g_K^{\rm CSS1}$ is the difference between an exact
quantity that is a function of $\Tsc{b}$ and the same quantity with
$\Tsc{b}$ replaced by $\bstarsc$.  We use this to get equality of the
separate terms on the two sides of Eq.\ 
\eqref{eq:CSS1.v.CSS2.Q.b}, which has the structure
\begin{equation}
\label{eq:XY}
X(\Tsc{b}) + \frac{\bstarsc^2}{\Tsc{b}^2} Y(\bstarsc) 
= X'(\Tsc{b}) +  \frac{\bstarsc^2}{\Tsc{b}^2}  Y'(\bstarsc) ,
\end{equation}
where we have segregated functions with the arguments $\Tsc{b}$ and
$\bstarsc$.  Each pair $(X,X')$ and $(Y,Y')$ represents corresponding
functions in the two schemes. Furthermore $X(\Tsc{b})$ is defined to
be $Y(\Tsc{b})-\frac{\bstarsc^2}{\Tsc{b}^2} Y(\bstarsc)$, and similarly
for $X'$, i.e., each is the difference between
an exact quantity at argument $\Tsc{b}$ and the same quantity at
argument $\bstarsc$.  Setting $\bmax=\infty$ gives 
$Y(\Tsc{b}) = Y'(\Tsc{b})$. It follows that
$Y(\bstarsc) = Y'(\bstarsc)$, and $X(\Tsc{b}) = X'(\Tsc{b})$.

Applying this to Eq.\ (\ref{eq:CSS1.v.CSS2.Q.b}) gives
\begin{align}
\label{eq:A.CSS1}
    A_{\rm CSS1}(a_s(\mubstar); C_1 )
  = {}& 
    - \frac{ \diff{ \tilde{K}(\bstarsc;\mubstar) } }
           { \diff{\ln\bstarsc^2} }
     + \frac{1}{2} \gamma_K(a_s(\mubstar))
 = 
     \left.
          - \frac{ \partial{ \tilde{K}(\bstarsc;\mu) } }
                 { \partial{\ln\bstarsc^2} }  
     \right|_{\mu \mapsto \mubstar} ,
\displaybreak[0]\\
    \frac{ \diff{g^{\rm CSS1}_K(\Tsc{b};\bmax) } }
           { \diff{\ln\Tsc{b}^2} } 
  = {}&
    \frac{ \diff{ g_K(\Tsc{b};\bmax) } }
           { \diff{\ln\Tsc{b}^2} } .
\end{align}

Next we substitute these results into Eq.\ (\ref{eq:CSS1.v.CSS2.Q}).
Again we equate the parts with the $g_K$ terms and the others to get
\begin{align}
\label{eq:B.CSS1}
    B_{\text{CSS1, DY}}(a_s(\muQ);C_1,C_2 )
  = {}&
    - \tilde{K}(C_1/\muQ;\muQ)
    - \gamma_j(a_s(\muQ)) 
    + \ln\frac{Q}{\muQ} \gamma_K(a_s(\muQ))
    - \frac{ \diff{ \ln H^{\rm DY}_{j\bar{\jmath}}(Q, \muQ, a_s(\muQ)) } }
           { \diff{\ln Q^2} }
\nonumber\\
  = {}&
    - \tilde{K}(C_1/\muQ;\muQ)
    - \frac{ \partial{ \ln H^{\rm DY}_{j\bar{\jmath}}(Q, \muQ, a_s(\muQ)) } }
           { \partial{\ln Q^2} } ,
\displaybreak[0]\\
\label{eq:gK.CSS1}
      g^{\rm CSS1}_K(\Tsc{b};\bmax) 
    ={}&
      g_K(\Tsc{b};\bmax).
\end{align}
Hence the ``non-perturbative'' $g_K$ function is the same in the two
formalisms, and the $A$ and $B$ functions are related to
perturbative quantities in the new formalism.  Calculations of
$A_{\rm CSS1}$ and $B_{\text{CSS1, DY}}$ were done to order $a_s^2$
by \citet{Davies:1984hs}, starting from
calculations of the $\Tsc{q}$-dependent Drell-Yan  cross section in
collinear factorization.
In the new formalism, instead of 2 quantities, there are 4
quantities to be determined: $H^{\rm DY}$, $\tilde{K}$, $\gamma_j$,
and $\gamma_K$.  To obtain them, we will supplement the existing
results for $A_{\rm CSS1}$ and $B_{\text{CSS1, DY}}$  by the results of other
calculations. These we will obtain in
Sec.\ \ref{sec:ff} from existing calculations of the quark form
factor.  A consistency condition will also be checked there.

Finally, we return to Eq.\ (\ref{eq:CSS1.v.CSS2}).  We substitute
into it the values for $A_{\rm CSS1}$, $B_{\text{CSS1, DY}}$, and 
$g^{\rm CSS1}_K$, etc.  
\begin{align}
  &\hspace*{-15mm} 
   e_j^2 \sum_{j_A} \int_{x_A}^1 \frac{ \diff{\xi_A} }{ \xi_A }
       f_{j_A/A}(\xi_A;\mubstar) 
  ~ \tilde{C}^{\text{CSS1, DY}}_{j/j_A}\xleft( \frac{x_A}{\xi_A},\bstarsc;
              \mubstar^2, \mubstar, C_2, a_s(\mubstar) \right)
  \times \mbox{Similar for hadron $B$}
\nonumber\\
  ={}&
  \sum_{j_A} \int_{x_A}^1 \frac{ \diff{\xi_A} }{ \xi_A }
       f_{j_A/A}(\xi_A;\mubstar) 
  ~ \tilde{C}^{\rm PDF}_{j/j_A}\xleft( \frac{x_A}{\xi_A},\bstarsc;
  \mubstar^2, \mubstar, a_s(\mubstar) \right)
  \times \mbox{Similar for hadron $B$}
\nonumber\\& \times
  H^{\rm DY}_{j\bar{\jmath}}(\mubstar/C_2, \mubstar, a_s(\mubstar))
  \exp\xleft[
        -2 \tilde{K}(\bstarsc;\mubstar) \ln C_2
      \right]
\nonumber\\& \times
  \exp\xleft[
        - g_{j/A}(x_A,\Tsc{b};\bmax)
        - g_{\bar{\jmath}/B}(x_B,\Tsc{b};\bmax)
        + g^{\rm CSS1}_{j/A}(x_A,\Tsc{b};\bmax)
        + g^{\rm CSS1}_{\bar{\jmath}/B}(x_B,\Tsc{b};\bmax) \label{eq:ABequality}
      \right] .
\end{align}
The same argument as was used for Eq.\
  (\ref{eq:CSS1.v.CSS2.Q.b}) applies here and shows that we have
  equality separately for the factors depending on $\bstarsc$ and the
  factors that involve the $g$ functions. Hence
\begin{equation}\
\label{eq:summed.g.1v2}
     g_{j/A}(x_A,\Tsc{b};\bmax)
     + g_{\bar{\jmath}/B}(x_B,\Tsc{b};\bmax)
  =  g^{\rm CSS1}_{j/A}(x_A,\Tsc{b};\bmax)
     + g^{\rm CSS1}_{\bar{\jmath}/B}(x_B,\Tsc{b};\bmax).
\end{equation}
To derive the corresponding relation for the individual
functions, we observe that each function is obtained from the
corresponding TMD parton density.  Now, the charge conjugation invariance of
QCD shows that an antiquark distribution in an antiparticle equals
the corresponding quark distribution in a particle, i.e.,
$f_{\bar{\jmath}/\bar{A}} = f_{j/A}$, and it follows that the same
relation applies to the $g$ functions. So by setting $B=\bar{A}$ and
$x_A=x_B$ in Eq.\ (\ref{eq:summed.g.1v2}), we obtain equality of the
individual $g$ functions, Eq.\ (\ref{eq:C.CSS1b}) below.

For the rest, we can factor out the collinear parton
densities,\footnote{To show this formally, one can take Mellin
  transforms in $x_A$ and $x_B$ to convert the convolutions to
  products.  Then one can use a set of hadrons of different flavors,
  and therefore with different ratios of different flavors of parton.}
to obtain
\begin{align}
   e_j^2 \tilde{C}^{\text{CSS1, DY}}_{j/j_A}\xleft( 
        \frac{x_A}{\xi_A},\bstarsc; \mubstar^2, \mubstar, C_2, a_s(\mubstar)
        \right)
  \times &
  \tilde{C}^{\text{CSS1, DY}}_{\bar{\jmath}/j_B}\xleft(
     \frac{x_B}{\xi_B},\bstarsc; \mubstar^2, \mubstar, C_2, a_s(\mubstar)
  \right)
\nonumber\\
  ={}&
  \tilde{C}^{\rm PDF}_{j/j_A}\xleft( 
        \frac{x_A}{\xi_A},\bstarsc;
        \mubstar^2, \mubstar, a_s(\mubstar)
  \right)
  \times 
  \tilde{C}^{\rm PDF}_{\bar{\jmath}/j_B}\xleft(
     \frac{x_B}{\xi_B},\bstarsc; \mubstar^2, \mubstar, a_s(\mubstar)
  \right)
\nonumber\\ & \times
  H^{\rm DY}_{j\bar{\jmath}}(\mubstar/C_2, \mubstar, a_s(\mubstar))
  \exp\xleft[
        -2 \tilde{K}(\bstarsc;\mubstar) \ln C_2
      \right].
\end{align}
This equation by itself does not determine how much of the
$H^{\rm DY}$ and the exponential factors is to be put with the
factor involving the quark $j$ and how much with the factor
involving the antiquark $\bar{\jmath}$. Again we appeal to charge
conjugation invariance in QCD, now to obtain charge-conjugation
relationships for the $\tilde{C}$ functions.  It follows that the
$H^{\rm DY}$ and the exponential factors must be assigned in equal
amounts to each $\tilde{C}$ coefficient.  Hence
\begin{align}
\label{eq:C.CSS1}
& \hspace*{-30mm}
 |e_j|
 \tilde{C}^{\text{CSS1, DY}}_{j/k}\xleft( \frac{x}{\xi},\bstarsc; \mubstar^2, \mubstar, 
 C_2, a_s(\mubstar) \right)
\nonumber \\
={}&
 \tilde{C}^{\rm PDF}_{j/k}\xleft( \frac{x}{\xi},\bstarsc;
  \mubstar^2, \mubstar, a_s(\mubstar) \right)
  \sqrt{H^{\rm DY}_{j\bar{\jmath}}(\mubstar/C_2, \mubstar, a_s(\mubstar))}
  \exp\xleft[
        - \tilde{K}(\bstarsc;\mubstar) \ln C_2
      \right],
\\
\label{eq:C.CSS1b}
   g^{\rm CSS1}_{j/H}(x,\Tsc{b};\bmax)
   ={}& g_{j/H}(x,\Tsc{b};\bmax).
\end{align}

Equations (\ref{eq:A.CSS1}), (\ref{eq:B.CSS1}),
(\ref{eq:gK.CSS1}) (\ref{eq:C.CSS1}), and (\ref{eq:C.CSS1b}) give
functions in the CSS1 formalism in terms of functions in the new
formalism.  We will see in Sec.\ \ref{sec:CSS2.calcs}, how to go in
the reverse direction, to obtain $\tilde{K}$ and $\tilde{C}$ in the
new formalism from functions in the CSS1 formalism. 

\end{widetext}

\subsection{Process dependence}

The above formulas give the relations between quantities in the
original CSS formula (\ref{eq:CSS.A.B}) for Drell-Yan
and those in the new TMD
factorization.  Most of the quantities in the new formalism are
process independent, because they concern properties of the TMD
functions.  These universal quantities are $\tilde{K}$, $\tilde{C}$,
$\gamma_j$, and $\gamma_K$, as well as $g_K$ and $g_{j/A}$.
Process dependence is confined to the hard scattering factor $\tilde{H}$, which 
would be $\tilde{H}^{\rm SIDIS}$ for SIDIS, and to the sign reversals between DY 
and SIDIS of the polarization-dependent TMD parton densities that are
time-reversal odd \cite{Collins:2002kn}.

In addition, for SIDIS we need the generally different $\tilde{C}$
functions for TMD fragmentation, the separately fitted functions
$g_{A/j}$ functions for the large $\T{b}$ behavior of fragmentation
functions, together with the $\T{b}$ dependence of other TMD
functions used in polarization-dependent processes.  

\section{TMD functions from fits with CSS1}
\label{sec:TMD.CSS1.fits}

Fits such as those of Refs.\ \cite{Landry:2002ix, Konychev:2005iy}
were given as results for the functions $g_K$ and $g_{j/H}$, but were
not presented in terms of actual TMD parton densities.  See also 
Refs.~\cite{Qiu:2000hf,Qiu:2002mu} which used a slightly different method for dealing with non-perturbative 
behavior at large transverse sizes. In this
section we show how to calculate the evolved TMD parton densities in
terms of the results of the fits.  We use the CSS2 definitions of the
TMD densities.

One advantage of expressing the results in terms of actual TMD
densities is that it facilitates comparison between different work.
For example, much recent phenomenological work
particularly for SIDIS,
e.g., \cite{Signori:2013mda},
works directly with TMD densities.  By contrast, 
the Drell-Yan fits in Refs.\ \cite{Landry:2002ix, Konychev:2005iy}
give results in terms of the TMD factorization formula in the
particular CSS1 form given in Eq.\ (\ref{eq:CSS.A.B}).  Other work might
use different forms and approximations for TMD factorization. The
genuine differences can be most directly assessed by comparison of
fitted results at level of the TMD parton densities.  It also provides
an invariant method of comparing the results of fits with different
values of $\bmax$.

Results of fits can then be presented in terms of evolved TMD
densities.  Then another advantage appears, that predictions for cross
sections can be made using the simple formula (\ref{eq:fact.CSS2}).
This differs from the elementary parton-model formula only by using
evolved TMD densities and by having higher-order corrections in the
hard scattering.  The higher-order corrections to the hard scattering
are suppressed by powers of $a_s(Q)$.

From results summarized in \cite{Collins:2014jpa}, we find
\cite{Aybat:2011zv} that the TMD parton densities are
\begin{widetext}
\begin{align}
\label{eq:TMD.pdf.from.fit}
  \tilde{f}_{j/H}(x,\T{b};Q^2,\muQ) 
 ={}&
  \exp\xleft[-g_{j/A}(x_A,\Tsc{b};\bmax) 
       - g_K(\Tsc{b};\bmax) \ln \frac{ Q }{ Q_0 } 
  \right]
\nonumber\\& \times
  \exp\xleft\{ 
       \tilde{K}(\bstarsc;\mubstar) \ln \frac{ Q }{ \mubstar }
       + \int_{\mubstar}^{\muQ}  \frac{ \diff{\mu'} }{ \mu' }
          \left[ \gamma_j(a_s(\mu')) 
                 - \ln\frac{Q}{ \mu' } \gamma_K(a_s(\mu'))
          \right]
  \right\}
\nonumber\\& \times
  \sum_{j_A} \int_{x_A}^1 \frac{ \diff{\xi} }{ \xi }
       f_{j_A/H}(\xi;\mubstar) 
  ~ \tilde{C}^{\rm PDF}_{j/j_A}\xleft( \frac{x}{\xi},\bstarsc;
                            \mubstar^2, \mubstar, a_s(\mubstar)
                    \right).
\end{align}
\end{widetext}

Note that when applying this formula to fits, one should be aware of
the issues raised in Ref.\ \cite{Collins:2014jpa}.  Among these are
that fits like those in Refs.\ \cite{Landry:2002ix, Konychev:2005iy}
used a quadratic form for the $\Tsc{b}$ dependence of the $g$
functions. However, the fits only determine the values of
these functions in a certain moderate range of $\Tsc{b}$.  When one
wants to use the results at lower $Q$ than the data in the
  fits, there is sensitivity to 
larger values of $\Tsc{b}$. Thus a simple extrapolation of a fitted
quadratic form may be quite inaccurate.

In addition, when the functions $g_K$ and $g_{j/A}$ are
  obtained by fitting data to factorization formulas like
  \eqref{eq:CSS.A.B}, the perturbative quantities, including those in
  the exponential, are calculated with truncated perturbation theory.
  The truncation errors then propagate to errors on the fitted
  functions compared with their true values, as strictly defined, for
  example, by Eqs.\ (13.60) and (13.68) of \cite{Collins:2011qcdbook}.
  Since the organization of the perturbative parts of TMD
  factorization differs between CSS1 and CSS2, the equalities of the
  ``non-perturbative'' functions in the two schemes is up to the
  effects of perturbative truncation errors.

\section{Obtaining the coefficients for CSS2}
\label{sec:CSS2.calcs}

The transformation to the $A,B,C$ form in Eq.\ (\ref{eq:CSS.A.B})
eliminated both the hard and soft factors present in the underlying
TMD factorization formula.  This enables the perturbative values of
the coefficients to be obtained from perturbative calculations of 
large $\Tsc{q}{}$ behavior in
collinear factorization instead of from separate calculations in the
TMD framework.  Taken to leading power in $\Tsc{q}/Q$, the collinear
hard scattering coefficients are matched to corresponding quantities
obtained from the perturbative expansion of Eq.\ (\ref{eq:CSS.A.B}).

The reason that this works is that there is a common domain of
validity of TMD factorization and collinear factorization at
intermediate transverse momentum, when $M \ll \Tsc{q} \ll Q$, where
$M$ is a typical hadronic scale.

Alternatively, direct calculations can be made in TMD
factorization with the use of
the definitions of the quantities involved---see \cite[Ch.\
13]{Collins:2011qcdbook}, \cite{Gehrmann:2012ze, Gehrmann:2014yya,
  Korchemsky:1987wg, Grozin:2014hna, Grozin:2015kna, Li:2016ctv} for
some examples at one, two, and three loops.

However, when using the first method, it is not sufficient
simply to match TMD and collinear
factorization.  It can be seen from formulae in Sec.\ \ref{sec:new.old}, that
separate knowledge of $H^{\rm DY}$, $\gamma_{j}$ and $\gamma_K$ is needed
as well.  After those values are obtained, which we will do,
the quantities $\tilde{K}$ 
and $\tilde{C}$ can be derived from the values of $A$, $B$, and $C$ in
the CSS1 scheme and hence from calculations of
  large-$\Tsc{q}{}$ behavior in collinear factorization.  
Observe that Eqs.\ (\ref{eq:B.CSS1}) and (\ref{eq:C.CSS1}) determine
$\tilde{K}$ and $\tilde{C}$ for particular values of their $\mu$ and
$\zeta$ arguments. Then evolution equations determine these
functions for general values of their arguments.
Since the values of 5 quantities are obtained
  from calculations of 6 quantities, one consistency condition also
  applies, which we can choose to be Eq.\ (\ref{eq:A.CSS1}).

The quantities, $H^{\rm DY}$, $\gamma_j$ and $\gamma_K$,
can be obtained from existing calculations of
the quark electromagnetic form factor, which have been done up to
3-loop order by \citet{Moch:2005id, Baikov:2009bg,
 Gehrmann:2010ue, Lee:2010cga}.  We will give a detailed
derivation of how to use these calculations in Sec.\ \ref{sec:ff}.  Then we
present the anomalous dimensions at order $a_s^3$ and the hard
coefficient and matching coefficients at order $a_s^2$ in Sec.\
\ref{sec:DY.values}, confirming results in Refs.\
\cite{Becher:2006mr}.

The reasons (already alluded to in \cite{Moch:2005id}) for
the success of this procedure are that
\begin{enumerate}

\item The quark form factor obeys factorization and evolution
  properties of a similar structure to that of TMD factorization
  \cite{Magnea:1990zb,Magnea:2000ss}.

\item The hard factor for DY (and SIDIS) is obtained from the
  same graphs as for the quark form factor, with subtractions
  of soft and collinear contributions.  So the DY hard factor is just
  the absolute value squared of the hard factor for the
   corresponding time-like quark form factor\footnote{We use
      ``Sud'' to denote ``Sudakov'', after the originator of work on
      the asymptotics of such form factors.  The hard scattering for
      SIDIS is, naturally, also the square of a quark form factor, but
      with space-like kinematics for the virtual photon.}:
    $H^{\text{DY}}_{j\bar{\jmath}} = \left|H^{\text{Sud,
        TL}}_{j\bar{\jmath}}\right|^2$. Unlike the hard scattering factor in
    collinear factorization, there is no contribution to $H^{\rm DY}$
    from graphs with emission of real partons. At leading power, the
    effects of real-emission graphs are only in the
    $\tilde{C}$-coefficients and in the
    $Y$-term.\footnote{The $Y$ term was defined in
          Ref.\ \cite{Collins:1981va, Collins:1984kg} as an additive
          correction to the TMD factorization term.  It implements
          matched asymptotic expansions for small and large $\Tsc{q}$,
          and thereby gives a result that agrees with large $\Tsc{q}{}$, 
          fixed-order collinear
          factorization at large $\Tsc{q}$ and TMD factorization at
          small $\Tsc{q}$.}

\item The bare collinear factors in a massless theory are
    scale-free and hence zero.

\item The anomalous dimensions $\gamma_j$ and $\gamma_K$ are
     related between the Drell-Yan process and
     the form factor.  They are also the same for parton densities
     in SIDIS and for fragmentation functions.  We will give the
     derivations later.

\item The extraction of the hard factor from the full form factor in
  massless QCD is made quite elementary because the massless integrals
  for its bare soft and collinear factors are scale free and hence
  zero. 

\end{enumerate}

In providing a complete treatment, we find some
complications in the case of the form 
factor concerning the phases of the collinear factors in relation to
the directions of the Wilson lines used in the definitions.  Some of
our results appear to be new, although they are closely related to
results by Magnea and Sterman \cite{Magnea:1990zb,Magnea:2000ss}.
To avoid interrupting the main flow of the argument, some of
the derivations are postponed to App.\ \ref{app:ff}.

\section{Analysis of the quark form factor}
\label{sec:ff}

\begin{figure*}
  \centering
  \begin{tabular}{c@{\hspace*{2cm}}c}
     \includegraphics[scale=0.5]{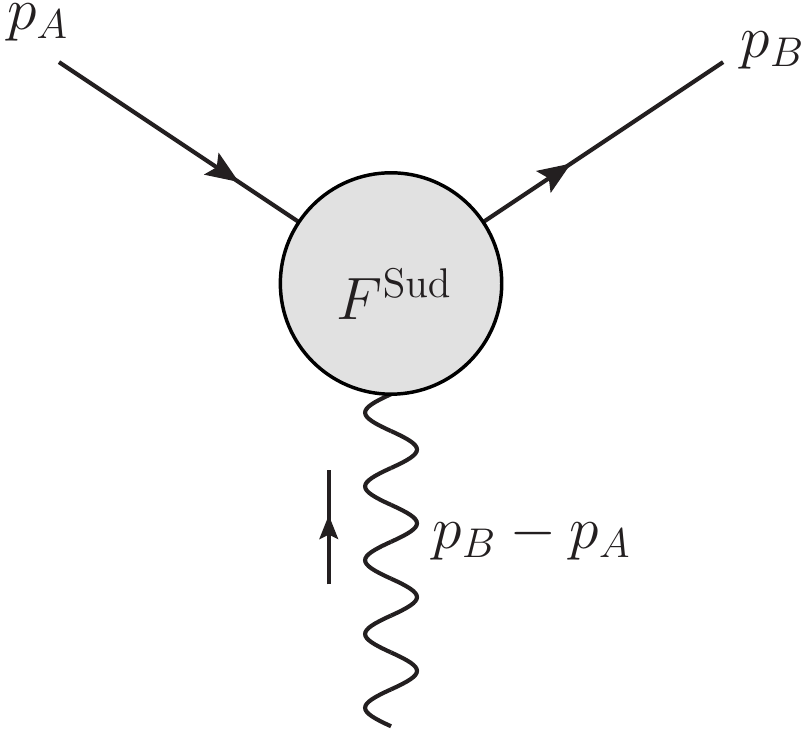}
  &
     \raisebox{2mm}{\includegraphics[scale=0.5]{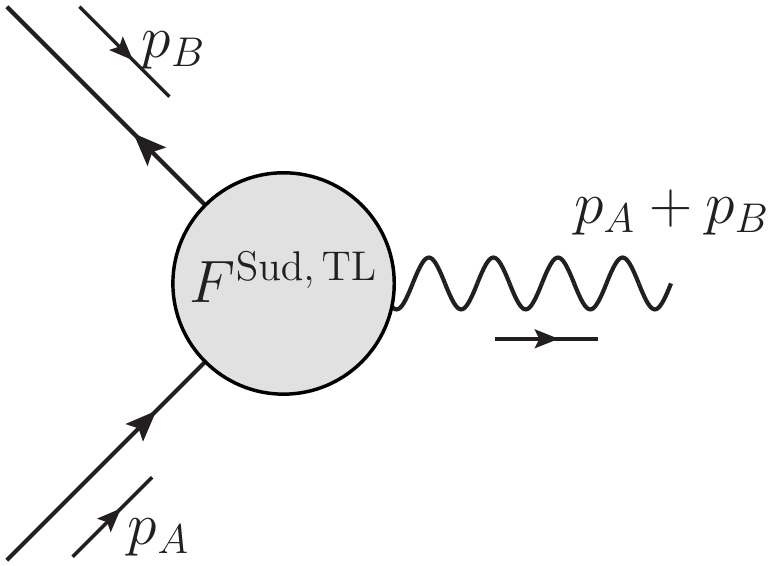}}
  \\ \vspace{1mm} \\
     (a) & (b)
  \end{tabular}  
  \caption{Graphs for the space-like (a) and time-like (b) quark form
    factor.}
  \label{fig:formfactors}
\end{figure*}

Let $F^{\text{Sud}}_j(Q_E^2)$ be the quark form factor,
defined in App.\ \ref{app:ff},
for the space-like electromagnetic
process $\gamma^*(q) + q_j(p_A) \to q_j(p_B)$ on a quark of flavor
$j$. It is illustrated in Fig.~\ref{fig:formfactors}(a).
The momentum transfer is $Q_E^2 = -q^2=-(p_B-p_A)^2$, normalized
to be positive for a space-like virtual photon.  The form-factor
for the space-like process is purely real.  It is normalized so that
its lowest-order term is $1$.  That is, a factor $e_j$ has
been divided out, where $e_j$ is the charge of the quark.

Results for the Drell-Yan process are obtained from the form factor
for the time-like process, $q_j(p_A) + \bar{q}_{\bar{\jmath}}(p_B) \to
\gamma^*(q)$ [Fig.~\ref{fig:formfactors}(b)]. The time-like form
factor is obtained by analytic continuation of the space-like form
factor to $Q_E^2=-Q^2-i\epsilon= -(p_A+p_B)^2-i\epsilon$, to give
$F^{\text{Sud, TL}}_{j\bar{\jmath}}(Q^2) =
F^{\text{Sud}}_j(-Q^2-i\epsilon)$.

\subsection{Factorization for form factor}
\label{sec:form.factor.factorization}

Factorization for the form factor was treated in massless QCD in
\cite{Magnea:1990zb,Magnea:2000ss,Moch:2005id}.  We will use the
specific formulation given by Collins in Ref.\
\cite{Collins:2011qcdbook}, with its definition of collinear factors
in terms of an unsubtracted ``collinear'' matrix
element and a combination of soft factors, with the relevant operators containing 
Wilson lines in particular directions.
\citet{Collins:2011qcdbook} gave results for the
form factor in the case of a massive Abelian theory, but using methods
later in \cite{Collins:2011qcdbook}, the results can be seen to generalize to massless
QCD, with results generally compatible with those of
\cite{Magnea:1990zb,Magnea:2000ss,Moch:2005id}.
In this section, we will mostly use only the massless case,
since that will be what is relevant for our calculations.

First we specify our conventions for how results are presented in
terms of coupling dependence, and for our use of the \MSbar{} scheme. 
Renormalized quantities are written
in terms of the coupling parameter $a_s$ defined in Eq.\
(\ref{eq:as.def}).
The bare coupling,\footnote{Although we generally follow the
conventions of \citet{Moch:2005id}, they use ``bare coupling'' to
refer to a differently normalized quantity than we do.} such as is
used in the Lagrangian, 
has the form
\begin{equation}
\label{eq:as0}
  a_{s,0} = \frac{ \mu^{2\epsilon} }{ S_\epsilon }  a_s
           \left( 1 + \sum_{n=1}^\infty \sum_{m=1}^n
                      A_{nm}\frac{ a_s^n }{ \epsilon^m }
           \right) ,
\end{equation}
where the space-time dimension is $n=4-2\epsilon$, and
\begin{equation}
  S_\epsilon = (4\pi e^{-\gammae})^{\epsilon}.
\end{equation}
The \MSbar{} scheme for coupling renormalization is defined
by the requirement that the 
renormalization counterterms have the form shown in (\ref{eq:as0}),
where there is an overall factor 
$\mu^{2\epsilon}/S_\epsilon$, and
there is otherwise a series of only negative powers of $\epsilon$ to
make the counterterms.  
The conventions specified above for the \MSbar{} scheme
correspond to those used by \citet{Moch:2005id}.

The implementation of \MSbar{} in Ref.\ \cite{Collins:2011qcdbook}
differed in two ways.  First there was a change of variable to replace
$a_s$ by $a_sS_\epsilon$; this does not change the renormalized
coupling at $\epsilon=0$ and so does not affect finite renormalized
quantities at the physical space-time dimension. Second, the value of
$S_\epsilon$ was changed to \cite{Collins:2011qcdbook}:
\begin{equation}
\label{eq:S.eps.JCC}
  S^{\rm JCC}_\epsilon = \frac{ (4\pi)^{\epsilon} }{ \Gamma(1-\epsilon) }.
\end{equation}
This has an advantage for the presentation of quantities whose
counterterms have 2 poles per loop.  The use of the form
(\ref{eq:S.eps.JCC}) for
$S_\epsilon$ amounts to a change of scheme for such quantities.  But
it is currently less standard, so our main results will use the
standard form.

For the calculations to be presented here, we will work within
  pure perturbation theory in strictly massless QCD.  Then the
  power-suppressed corrections, such as we notated in earlier statements
  of TMD factorization, are zero.
Factorization for the time-like form factor in massless QCD has the
form
\begin{multline}
\label{eq:fact.sud.TL}
  \hspace*{-3mm}
  F^{\text{Sud}}_j\xleft( \frac{-Q^2-i\epsilon}{\mu^2}; a_s(\mu),\epsilon \right)
\\ 
= H^{\text{Sud}}_j \xleft( \frac{-Q^2-i\epsilon}{\mu^2}; a_s(\mu), \epsilon \right)
  \left[ C_j^{\rm Sud}(Q^2,\mu, a_s(\mu),\epsilon)
  \right]^2,
\end{multline}
in the notation of App.\ \ref{app:ff}.  Here
$H^{\text{Sud}}_j$ is the hard factor, finite as
$\epsilon\to0$, with subtractions for all collinear and soft
contributions.  One of the collinear factors $C_j$ is for the quark of
flavor $j$.  Its first argument is the CSS $\zeta$ argument set to the
value $Q^2$.  The second collinear factor is for the antiquark, and by
charge-conjugation invariance it equals the quark's collinear factor.
By use of the operator definitions given in \cite[Ch.\
10]{Collins:2011qcdbook}, the collinear factors include to leading
power not only all contributions from collinear momenta but all soft
contributions as well.  To achieve this correctly, the Wilson lines
used in the operator matrix elements used to define $C_j$ must be past
pointing when the quark and antiquark are incoming
\cite{Collins:2011qcdbook}.

We will also use factorization for the space-like case.  By results in
App.\ \ref{app:ff}, one of the collinear factors must be complex
conjugated, so that we have:
\begin{multline}
\label{eq:fact.sud.SL}
\hspace*{-3mm}
  F^{\text{Sud}}_j(Q_E^2/\mu^2; a_s(\mu),\epsilon)
\\ \hspace*{-5mm}
= H^{\text{Sud}}_j(Q_E^2/\mu^2; a_s(\mu), \epsilon)
  \left| C^{\rm Sud}_j(Q_{E}^2,\mu, a_s(\mu),\epsilon)
  \right|^2,
\end{multline}
with $Q_E^2$ being positive for
the space-like case. Both of $F$ and $H$ are now real.

In each of Eqs.\ (\ref{eq:fact.sud.TL}) and (\ref{eq:fact.sud.SL}),
the different factors on the right-hand side depend on the same
variables, so at first sight 
there might appear to be no content.  The significance of
factorization is from the segregation of contributions from different
regions of momenta. The lack of collinear and soft contributions to
$H^{\rm Sud}$ imply that it has no divergences and also has no large logarithms
when $\mu$ is of order $Q$; then it can be predicted perturbatively
when $Q$ is large enough.  The collinear factors have collinear and
soft contributions, and they diverge in the massless limit. Furthermore, their
definition allows useful equations to be derived for both their $\mu$
and $Q$ dependence. If masses were restored, then Eqs.\
(\ref{eq:fact.sud.TL}) and (\ref{eq:fact.sud.SL}) would be true to
leading power in masses divided by $Q$ for large $Q$, and the
collinear factors would be mass-dependent\footnote{If all fields were
  massive, then the collinear factors no longer have actual collinear
  and soft divergences, of course.}, but the hard factor would remain
mass-independent with an unchanged value.

We will use evolution equations in the form found\footnote{See also Refs.\
\cite{Magnea:1990zb,Magnea:2000ss}.} in Ref.\ \cite{Collins:2011qcdbook}.  In  
addition, we will need extra results derived 
in App.\ \ref{app:ff} concerning the real and imaginary parts of the
anomalous dimensions; these will be important in relating anomalous
dimensions for the form factor to anomalous dimensions for the
Drell-Yan process.

The renormalization-group (RG) equation for the collinear factor is
\begin{align}
\label{eq:evo.C}
  \frac{\diff{ \ln C^{\rm Sud}_j}}{\diff{\ln \mu}}
\hspace*{-3mm}&
\nonumber\\
  = {}&
   \frac{1}{2} \gamma_j(a_s(\mu))
   + i \frac{\pi}{4}\gamma_K(a_s(\mu))
   - \frac{1}{4}\gamma_K(a_s(\mu)) \ln \frac{ Q^2 }{ \mu^2 }
\nonumber\\
  = {}&
   \frac{1}{2} \gamma_j(a_s(\mu))
   - \frac{1}{4}\gamma_K(a_s(\mu)) \ln \frac{ -Q^2-i\epsilon }{ \mu^2 }.
\end{align}
It is proved in App.\ \ref{app:ff} that the anomalous dimension functions $
\gamma_j$ and $\gamma_K$ are both real, and that the imaginary part on the 
right-hand side is as shown.  The normalizations of these functions are arranged 
so that they are exactly the same as the corresponding quantities in TMD 
factorization for the Drell-Yan process, with conventions as in 
Ref.\ \cite{Collins:2014jpa}. The equality of these quantities between the 
Drell-Yan cross section and the Sudakov form factor is because the anomalous 
dimensions are determined by the 
renormalization of the same virtual loops containing the same operators. Their 
contribution to the Drell-Yan cross section is obtained by the absolute value 
squared of the sum of graphs for the form factor.  
Thus $H^{\text{DY}}_{j\bar{\jmath}} =
|H^{\text{Sud}}_j((-Q^2-i\epsilon)/\mu^2)|^2$, while the anomalous
dimensions are $\gamma_j$ and $\gamma_K$, with cancellation of the imaginary 
part that appears in Eq.\ (\ref{eq:evo.C}).

Note that sometimes \cite{Collins:2011qcdbook}
$\gamma_j(a_s(\mu))$ is given a second argument, as in $\gamma_j(a_s(\mu);
\zeta/\mu^2)$.  The $\zeta$ dependence corresponds to the $Q^2$ dependence 
in Eq.\ (\ref{eq:evo.C}), and $\gamma_j(a_s(\mu))$ in Eqs.\ (\ref{eq:evo.C})
and (\ref{eq:evo.H}) corresponds to $\gamma_j(a_s(\mu);1)$ in the
  other notation.

The rapidity evolution equation for the collinear factor is
\begin{equation}
\label{eq:C.rap.evol}
  \frac{\partial C_j^{\rm Sud} }{\partial \ln Q}
  =
  \frac{1}{2} K^{\rm Sud}(a_s,\epsilon),
\end{equation}
with $K^{\rm Sud}$ obeying the RG equation
\begin{equation}
\label{eq:KsudRG}
 \frac{\diff{K^{\rm Sud}}}{\diff{\ln\mu}}
  = - \gamma_K(a_s)\, .
\end{equation}
Note that $K^{\rm Sud}$ has no explicit dependence on $Q$ and $\mu$;
it has soft divergences as $\epsilon\to0$, and would be finite (but mass 
dependent) in a massive theory or in a theory with confinement.

In the remainder of this section, we will work with the
  time-like form factor and hard part, using the notations
$F^{\text{Sud,~TL}}=F^{\text{Sud}}_j((-Q^2-i\epsilon)/\mu^2)$ and
$H^{\text{Sud,~TL}}=H^{\text{Sud}}_j((-Q^2-i\epsilon)/\mu^2)$.

Since the form factor is RG-invariant, it follows from Eqs.\ \eqref{eq:fact.sud.TL} 
and \eqref{eq:evo.C} that the RG equation for $H$ is
\begin{multline}
\label{eq:evo.H}
  \frac{\diff{\ln H^{\text{Sud,~TL}}}}{\diff{\ln \mu}}
\\
   =
    - \gamma_j(a_s(\mu))
    - i \frac{\pi}{2}\gamma_K(a_s(\mu))
    + \frac{1}{2} \gamma_K(a_s(\mu)) \ln \frac{ Q^2 }{ \mu^2 } .
\end{multline}

Each of the collinear factors in factorization (\ref{eq:fact.sud.TL})
is a bare collinear factor times an ultra-violet renormalization
factor.  It will be convenient to work with logarithms of the factors, for which
renormalization is additive.  We have
\begin{multline}
\label{eq:ln.F.SL}
    \ln F^{\rm Sud,~TL}(Q^2)
    = \ln H^{\rm Sud,~TL}
      + 2 \ln C_j^{\rm bare}
\\
      + D(a_s,\epsilon) - i\pi E(a_s,\epsilon)
      + \ln\frac{Q^2}{\mu^2} E(a_s,\epsilon) \, ,
\end{multline}
where the terms involving $E$ and $D$ implement counterterms
  for $\ln C_j$; the linearity in $\ln(Q^2/\mu^2)$ follows from Eq.\
(\ref{eq:C.rap.evol}), and the lack of $Q$ dependence of $K^{\rm Sud}$.
It is shown in App.\ \ref{app:ff} that each of
$D$ and $E$ is real, and that there is an imaginary
term $-i\pi E$, as in Eq.\ (\ref{eq:ln.F.SL}).  Each of $D$ and
$E$ has the usual \MSbar{} form: 
\begin{align}
\label{eq:Dexpand}
  D ={}& \sum_{n=1}^\infty \sum_{m=1}^{n+1}
               D_{nm}\frac{ a_s^n }{ \epsilon^m }
\\
\label{eq:Eexpand}
  E ={}& \sum_{n=1}^\infty \sum_{m=1}^n
               E_{nm}\frac{ a_s^n }{ \epsilon^m }\, .
\end{align}
That the highest powers of $1/\epsilon$ in each order are as shown
can be deduced from the evolution equations. 

In the massless case, all loop integrals for the unsubtracted bare collinear
factor are scale-free and hence zero
\cite{Magnea:1990zb,Magnea:2000ss}.  There remains only the lowest
order term, which is unity.  Hence, to all orders of perturbation
theory $\ln C^{\rm bare}_j=0$.  Therefore
\begin{multline}
\label{eq:ln.H}
\hspace*{-3mm}
  \ln H^{\rm Sud,~TL}(Q^2)
  =
\\
  \ln F^{\rm Sud,~TL}(Q^2)
   - D + i\pi E - \ln\frac{Q^2}{\mu^2} E
\quad
\text{(massless),}
\end{multline}
so that the finite quantity $\ln H^{\rm Sud}$ can be obtained from the
massless 
$\ln F^{\rm Sud}$ simply by subtracting \MSbar{} poles.  The poles initially
arise as ultra-violet counterterms.  But because of the zero value of
the scale-free integrals for the collinear factor, these counterterms
now subtract numerically opposite collinear and soft divergences in the logarithm 
of the form factor, $\ln F$.

We show in App.\ \ref{app:ff} that almost the same formula
  applies to the space-like case, with $F$ replaced by its space-like
  version, with omission of the imaginary term $i\pi E$ and with
  otherwise the same values of $D$ and $E$. It follows that the
  space-like hard part is obtained from the time-like hard part by the
  same analytic continuation that applies to the form factor
  itself.  We have already used this result in discussing
  Eq.~\eqref{eq:fact.sud.SL}.

Now in RG equations for the massless theory, the derivative of any
quantity $X$ with respect to $\mu$ is
\begin{equation}
  \frac{\diff{X}}{\diff{\ln\mu}}
  =  
  \frac{\partial X }{\partial \ln\mu }
  - 2 \left( 
          \epsilon a_s + \sum_{n=0}^\infty \beta_n
          a_s^{n+2}
      \right)
  \frac{\partial X }{\partial a_s },
\end{equation}
where the $\beta_n$ are the usual coefficients that control the
running of the coupling; they can obtained from the expansion of the
bare coupling in powers of the renormalized coupling.  In the
calculations in this paper we will only need the following terms:
\begin{equation}
\label{eq:as0.NLO}
  a_{s,0} = \frac{\mu^{2\epsilon}}{S_\epsilon} a_s
           \left[ 1 - \beta_0 \frac{ a_s }{ \epsilon }
                  + a_s^2 \left( \frac{\beta_0^2}{\epsilon^2} - \frac{\beta_1}{2\epsilon} \right)
                  + \dots
           \right] ,
\end{equation}
with the well-known values
\begin{subequations}
\begin{align}
  \beta_0 ={}& \frac{11}{3} C_A - \frac{2}{3} n_f,
\\
  \beta_1 ={}& \frac{34}{3} C_A^2 - \frac{10}{3} C_A n_f -2C_Fn_f.
\end{align}
\end{subequations}

The pole terms in the massless $\ln F^{\rm Sud,~TL}$ enable us to
deduce $D$ and $E$ (from Eq.\ (\ref{eq:ln.H}) given that $H^{\rm Sud}$
  is finite), and hence the anomalous dimensions.  The
calculation of the anomalous dimensions arise because the renormalized
collinear factor obeys
\begin{align}
  2 \ln C^{\rm Sud} ={}& 2 \ln C^{\rm bare} 
   + D -i\pi E + \ln\frac{Q^2}{\mu^2} E,
\nonumber\\
  ={}&  D -i\pi E + \ln\frac{Q^2}{\mu^2} E
       \qquad\mbox{(massless)},
\end{align}
and the bare quantity is RG invariant.  From \eqref{eq:evo.C} we find
\begin{widetext}
\begin{equation}
\label{eq:anomdims}
    \gamma_j - \frac12 \ln\frac{Q^2}{\mu^2} \gamma_K 
  =
    \frac{ \diff{(D+\ln\frac{Q^2}{\mu^2}E)} }{ \diff{\ln\mu} }
  =
     -2E
     - 2 \left(
         \epsilon a_s + \sum_{n=0}^\infty \beta_n
         a_s^{n+2}
       \right)
     \frac{\partial (D+\ln\frac{Q_E^2}{\mu^2}E) }{\partial a_s }\, ,
\end{equation}
\end{widetext}
Hence
\begin{equation}
    \gamma_K 
  =
     4 \left(
         \epsilon a_s + \sum_{n=0}^\infty \beta_n
         a_s^{n+2}
       \right)
     \frac{\partial E }{\partial a_s }
\label{eq:gammaKE}
\end{equation}
and
\begin{equation}
    \gamma_j
  =
     -2E 
     - 2 \left(
         \epsilon a_s + \sum_{n=0}^\infty \beta_n
         a_s^{n+2}
       \right)
     \frac{\partial D }{\partial a_s }.
\label{eq:gammajD}
\end{equation}

Let us define the expansions
\begin{align}
\gamma_K ={}& \sum_{n=1}^\infty \gamma_{K,n} a_s^n\, ,
\\
\gamma_j ={}& \sum_{n=1}^\infty \gamma_{j,n} a_s^n\, .
\end{align}
Matching terms on each side of Eqs.~\eqref{eq:gammaKE} and \eqref{eq:gammajD},
we deduce that the first two coefficients in
$\gamma_K$ are obtained from the single-pole counterterms in $E$:
\begin{subequations}
\label{eq:gammaK.1.2}
\begin{align}
\label{eq:gammaK.1}
  \gamma_{K,1} ={}& 4E_{1,1} ,
\\
\label{eq:gammaK.2}
  \gamma_{K,2} ={}& 8E_{2,1} .
\end{align}
\end{subequations}
The higher-pole counterterms are then determined since $\gamma_K$
has no $1/\epsilon$ poles. So all such poles must cancel
on the right side of Eqs.~\eqref{eq:gammaKE} and \eqref{eq:gammajD}. 
This gives
\begin{equation}
\label{eq:E.consist}
  E_{2,2} = -\frac{ \beta_0 \gamma_{K,1} }{ 8 }.
\end{equation}
Similar equations apply at higher orders, but we do not derive them
here.

Similarly, for $\gamma_j$ and $D$, we have
\begin{subequations}
\label{eq:gammaj.1.2}
\begin{align}
\label{eq:gammaj.1}
  \gamma_{j,1} ={}& -2 D_{1,1} ,
\\
\label{eq:gammaj.2}
  \gamma_{j,2} ={}& -4 D_{2,1} ,
\end{align}
\end{subequations}
and
\begin{subequations}
\label{eq:D.consist}
\begin{align}
  D_{1,2} ={}& -\frac{ \gamma_{K,1} }{ 4 } ,
\\
  D_{2,3} ={}& \frac{ 3\beta_0 \gamma_{K,1} }{ 16 } ,
\\
  D_{2,2} ={}& \frac{ \beta_0 \gamma_{j,1} }{ 4 }
              - \frac{ \gamma_{K,2} }{ 16 } .
\end{align}
\end{subequations}

\subsection{Coefficients for quark form factor}
\label{sec:Sud.from.MVV}

To obtain the actual values for the coefficients for the anomalous
dimensions and the hard factor, we start from results for the massless
form factor that were presented in Ref.\ \cite{Moch:2005id} as an
expansion in powers of the bare coupling:
\begin{equation}
\label{eq:Fn.def}
  F^{\rm Sud}
  = 1 +
    \sum_{n=1}^\infty \left( a_{s,0} S_\epsilon Q_E^{-2\epsilon} \right)^n
       \mathcal{F}_n(\epsilon),
\end{equation}
with $Q_E^2=-Q^2-i\epsilon$ for the time-like case that we need to obtain
results for the Drell-Yan process.  We then express the form factor in
terms of the renormalized coupling by Eq.\ (\ref{eq:as0}), of which we
will only need the two-loop expansion \eqref{eq:as0.NLO}.

We use Laurent expansions about $\epsilon=0$ of the coefficients
$\mathcal{F}_n(\epsilon)$ in Eq.\ (\ref{eq:Fn.def}), with the notation
\begin{equation}
\label{eq:Fnm.def}
  \mathcal{F}_n(\epsilon)
  = \sum_{m=-\infty}^{2n} \frac{F_{n,m}}{\epsilon^m}.
\end{equation}
That the highest power of $1/\epsilon$ is twice the number of loops
can be obtained from the evolution equations. For our calculations,
values for the relevant coefficients $F_{n,m}$ can be read off Eqs.\
(3.5) and (3.6) in \cite{Moch:2005id}.  

The logarithm of the form factor has
the following expansion in powers of the renormalized coupling
\begin{widetext}
\begin{equation}
\label{eq:ln.F.exp}
  \ln F^{\rm Sud}
  = a_s \left( \frac{Q_E^2}{\mu^2} \right)^{-\epsilon}
                 \mathcal{F}_1(\epsilon)
    + a_s^2
                 \left( \frac{Q_E^2}{\mu^2} \right)^{-2\epsilon}
                 \left[ \mathcal{F}_2(\epsilon) 
                        - \frac{\mathcal{F}_1(\epsilon)^2}{2}
                 \right]
    - a_s^2 \left( \frac{Q_E^2}{\mu^2} \right)^{-\epsilon}
                 \frac{ \beta_0 \mathcal{F}_1(\epsilon) }{ \epsilon} 
    + O(a_s^3) .
\end{equation}  
Now Eq.\ \eqref{eq:ln.H} shows that we can determine $D$ and $E$ from
the poles at $\epsilon=0$ in the coefficients in \eqref{eq:ln.F.exp},
and $\ln H$ from the finite remainder.  For $D$ and $E$ we
used a Mathematica program to obtain the following values up
to 3-loop order from coefficients in Ref.\ \cite{Moch:2005id}:
\begin{align}
  \label{eq:D}
  D ={}& 
    -a_s C_F
     \left(
     \frac{ 2 }{ \epsilon^2 }
     +\frac{ 3 }{ \epsilon }
     \right)
\nonumber\\&
    + a_s^2
      \Biggl\{
         C_F^2 
           \left[
              \frac{ - \frac34 + \pi^2 - 12 \zeta_3 }
                   { \epsilon }
           \right]
         + C_AC_F
             \left[
                   \frac{ \frac{11}{2} }{ \epsilon^3 }
                   + \frac{ \frac{16}{9} + \frac{\pi^2}{6} }
                          { \epsilon^2 }
                   + \frac{ - \frac{961}{108} - \frac{11\pi^2}{12} + 13 \zeta_3 }
                          { \epsilon }
           \right]
         + n_fC_F
           \left[
              - \frac{1}{\epsilon^3} 
              - \frac{ \frac{4}{9} }{ \epsilon^2 }
              + \frac{ \frac{65}{54} + \frac{\pi^2}{6} }{ \epsilon }
           \right]
      \Biggr\}
\nonumber\\&
   + a_s^3
     \Biggl\{
         C_F n_f^2 \left[ - \frac{44}{81 \epsilon^4} 
                      - \frac{8}{243 \epsilon^3}
                      + \frac{1}{\epsilon^2} \left( \frac{46 }{81}
                                            + \frac{2 \pi^2}{27}
                                    \right)
                      + \frac{1}{\epsilon} \left( \frac{2417 }{2187}
                                           - \frac{10 \pi^2 }{81}
                                           - \frac{8 \zeta_3}{81}
                                    \right)
               \right]
\nonumber\\&\hspace*{1cm}
        + C_F^2 n_f \left[ -\frac{16}{9 \epsilon^3} 
                       + \frac{1}{\epsilon^2} \left( - \frac{8}{27}
                                             + \frac{4 \pi^2}{9}
                                             - \frac{64 \zeta_3}{9}
                                      \right)
                       + \frac{1}{\epsilon} \left( \frac{2953}{162}
                                            - \frac{13 \pi^2}{27}
                                            - \frac{14 \pi^4}{81}
                                            + \frac{256 \zeta_3}{27}
                                     \right)
                 \right]
\nonumber\\&\hspace*{1cm}
        + C_A^2 C_F \left[ -\frac{1331}{81 \epsilon^4}
                        + \frac{1}{\epsilon^3} \left( \frac{2866}{243}
                                             - \frac{55 \pi^2}{81}
                                       \right)
                        + \frac{1}{\epsilon^2} \left( \frac{11669 }{486}
                                             + \frac{1625 \pi^2}{486}
                                             - \frac{22 \pi^4}{405}
                                             - \frac{902 \zeta_3}{27}
                                       \right)
\right.\nonumber\\&\hspace*{2.5cm}\left.
                        + \frac{1}{\epsilon} \left( - \frac{139345}{8748}
                                             - \frac{7163 \pi^2}{1458}
                                             - \frac{83 \pi^4}{270}
                                             + \frac{3526 \zeta_3}{27}
                                             - \frac{44 \pi^2 \zeta_3}{27}
                                             - \frac{136 \zeta_5}{3}
                                      \right)
                  \right]
\nonumber\\&\hspace*{1cm}
        + \frac{C_F^3}{\epsilon} \left( -\frac{29 }{6}
                              - \pi^2
                              - \frac{8\pi^4}{15}
                              - \frac{68 \zeta_3}{3}
                              + \frac{16 \pi^2 \zeta_3}{9}
                              + 80 \zeta_5
                       \right)
\nonumber\\&\hspace*{1cm}
        + C_AC_F n_f \left[ \frac{484 }{81 \epsilon^4}
                        + \frac{1}{\epsilon^3} \left( - \frac{752}{243}
                                              + \frac{10 \pi^2}{81}
                                       \right)
                        + \frac{1}{\epsilon^2} \left(
                                          - \frac{2068}{243}
                                          - \frac{238 \pi^2}{243}
                                          + \frac{212 \zeta_3}{27}
                                       \right)
\right.\nonumber\\&\hspace*{3cm}\left.
                        + \frac{1}{\epsilon} \left(
                                         - \frac{8659}{2187}
                                         + \frac{1297 \pi^2}{729}
                                         + \frac{11 \pi^4}{135}
                                         - \frac{964 \zeta_3}{81}
                                      \right)
                   \right]
\nonumber\\&\hspace*{1cm}
          + C_AC_F^2 \left[
                          \frac{1}{\epsilon^2} \left(
                                         \frac{11}{6}
                                         - \frac{22 \pi^2}{9}
                                         + \frac{88 \zeta_3}{3 \epsilon^2}
                                      \right)
                        + \frac{1}{\epsilon} \left(
                                         - \frac{151}{12}
                                         + \frac{205 \pi^2}{27}
                                         + \frac{247 \pi^4}{405}
                                         - \frac{844 \zeta_3}{9}
                                         - \frac{8\pi^2 \zeta_3}{9}
                                         - 40 \zeta_5
                                      \right)
                   \right]
     \Biggr\}
\nonumber\\&
   + O(a_s^4),
\displaybreak[0]\\
  \label{eq:E}
  E ={}&   
    a_s \frac{ 2 C_F  }{ \epsilon }
    + a_s^2
      \Biggl\{
         C_FC_A \left[
                 - \frac{11}{3\epsilon^2}
                 + \frac{1}{\epsilon} \left( \frac{67}{9} - \frac{\pi^2}{3} \right)
               \right]
         + C_Fn_f \left[
                  \frac{2}{3\epsilon^2}
                  - \frac{10}{9\epsilon}
                 \right]
      \Biggr\}
\nonumber\\&
    + a_s^3
      \Biggl\{
         C_F n_f^2 \left[ \frac{8}{27 \epsilon^3}
                      - \frac{40}{81 \epsilon^2}
                      - \frac{8}{81 \epsilon}
               \right]
         + C_A C_F n_f \left[ -\frac{88}{27 \epsilon^3}
                          + \frac{1}{\epsilon^2} \left( \frac{668}{81}
                                                - \frac{4 \pi^2}{27}
                                        \right)
                          + \frac{1}{\epsilon} \left( - \frac{418}{81}
                                               + \frac{40 \pi^2}{81}
                                               - \frac{56 \zeta_3}{9}
                                        \right)
                   \right]
\nonumber\\&\hspace*{1cm}
         + C_A^2 C_F \left[
                  \frac{242}{27 \epsilon^3}
                  + \frac{1}{\epsilon^2} \left( -\frac{2086}{81}
                                       + \frac{22 \pi^2}{27}
                                 \right)
                  +\frac{1}{\epsilon} \left( \frac{245}{9}
                                      - \frac{268 \pi^2}{81}
                                      + \frac{22 \pi^4}{135}
                                      + \frac{44 \zeta_3}{9}
                               \right)
                  \right]
\nonumber\\&\hspace*{1cm}
        + C_F^2 n_f \left[ \frac{4}{3 \epsilon^2}
                       + \frac{1}{\epsilon} \left( - \frac{55}{9}
                                            + \frac{16 \zeta_3}{3}
                                     \right)
                \right]
      \Biggr\}
\nonumber\\&
   + O(a_s^4).
\end{align}
%

The values for $\gamma_j$ and $\gamma_K$ are deduced from
Eqs.\ \eqref{eq:gammaKE} and \eqref{eq:gammajD}:
\begin{align}
  \label{eq:gammaj}
  \gamma_j ={}&   
    6 C_F a_s
    + a_s^2
      \left[
           C_F^2 \left( 3 - 4 \pi^2 + 48 \zeta_3 \right)
           + C_FC_A \left( \frac{961}{27} + \frac{11 \pi^2}{3} - 52 \zeta_3 \right)
           + C_Fn_f \left( - \frac{130}{27} - \frac{2\pi^2}{3} \right)
     \right]
\nonumber\\&
    + a_s^3
      \biggl[
         C_F^2 n_f \left( - \frac{2953}{27} + \frac{26 \pi^2}{9}
                      + \frac{28 \pi^4}{27} - \frac{512\zeta_3}{9}
               \right)
           +C_F n_f^2 \left( - \frac{4834}{729} + \frac{20 \pi^2}{27}
                         + \frac{16 \zeta_3}{27}
                  \right)
\nonumber\\&\hspace*{1cm}
           +C_F^3 \left( 29 + 6\pi^2 + \frac{16 \pi^4}{5}
                      + 136\zeta_3 - \frac{32 \pi^2 \zeta_3}{3}- 480 \zeta_5
               \right)
\nonumber\\&\hspace*{1cm}
            +C_A^2 C_F \left(\frac{139345}{1458}+\frac{7163 \pi^2}{243}+\frac{83 \pi^4}{45}-\frac{7052\zeta_3}{9}+\frac{88 \pi^2 \zeta_3}{9}+272 \zeta_5\right)
\nonumber\\&\hspace*{1cm}
            + C_AC_F n_f \left(\frac{17318}{729}-\frac{2594 \pi^2}{243}-\frac{22 \pi^4}{45}+\frac{1928\zeta_3}{27}\right)
\nonumber\\&\hspace*{1cm}
            + C_AC_F^2 \left(\frac{151}{2}-\frac{410 \pi^2}{9}-\frac{494 \pi^4}{135}+\frac{1688 \zeta_3}{3}+\frac{16 \pi^2 \zeta_3}{3}+240 \zeta_5\right)
      \biggr]
\nonumber\\&
   + O(a_s^4),
%
\\
  \label{eq:gammaK}
  \gamma_K ={}&   
    8 C_F a_s
    + a_s^2
      \left[
             C_A C_F \left( \frac{536}{9}- \frac{8\pi^2}{3} \right)
             - \frac{80}{9} C_F n_f
      \right]
\nonumber\\&
    + a_s^3
      \left[
          - \frac{32}{27} C_F n_f^2
          + C_A C_F n_f \left(
                      - \frac{1672}{27} + \frac{160 \pi^2}{27}
                      - \frac{224 \zeta_3}{3}
                    \right)
          + C_A^2 C_F \left( \frac{980}{3}
                          - \frac{1072 \pi^2}{27}
                          + \frac{88\pi^4}{45}
                          + \frac{176 \zeta_3}{3}
                   \right)
\right.\nonumber\\&\hspace*{1cm}\left.
          + C_F^2 n_f \left( - \frac{220}{3} + 64 \zeta_3 \right)
      \right]
\nonumber\\&
   + O(a_s^4).
\end{align}
%
We also verify that the consistency conditions \eqref{eq:E.consist}
and \eqref{eq:D.consist} are obeyed. The above values are in agreement
with the results of \citet{Gehrmann:2010ue}, after allowing for the
different normalizations of their anomalous dimensions.

Finally, the Sudakov hard factor at $\epsilon=0$ is
\begin{align}
  \label{eq:Hsud}
  H^{\text{Sud}} ={}&   
      1
      + C_F a_s
        \left( - 8 + \frac{\pi^2}{6} + 3t - t^2 \right)
\nonumber\\&\hspace*{-10mm}
      + a_s^2 \Biggl\{
        C_F^2 \left[
                   \frac{255}{8}
                   + \frac{7\pi^2}{2}
                   - \frac{83 \pi^4}{360}  
                   - 30 \zeta_3
                   + t \left( - \frac{45}{2} - \frac{3\pi^2}{2} + 24\zeta_3 \right)
                   + t^2 \left( \frac{25}{2} - \frac{\pi^2}{6} \right)
                   - 3 t^3
                   + \frac12 t^4
        \right]
\nonumber\\& \hspace*{0mm}
        +C_FC_A\Biggl[
                 - \frac{51157}{648}
                 - \frac{337 \pi^2}{108}
                 + \frac{11 \pi^4}{45}
                 + \frac{313}{9} \zeta_3
                 + t \left( \frac{2545}{54}
                            + \frac{11 \pi^2}{9} - 26 \zeta_3
                      \right)
                 + t^2 \left( -\frac{233}{18} + \frac{\pi^2}{3} \right)
                 + \frac{11}{9} t^3
        \Biggr]
\nonumber\\& \hspace*{0mm}
        +C_Fn_f\Biggl[
                  \frac{4085}{324} + \frac{23 \pi^2}{54} + \frac{2}{9} \zeta_3
                  + t \left( - \frac{209}{27} - \frac{2 \pi^2}{9} \right)
                  + \frac{19}{9} t^2 - \frac{2}{9} t^3
        \Biggr]
      \Biggr\}
\nonumber \\ &{}\hspace*{-10mm}
 +a_s^3 \Biggl\{ C_F^3 \Biggl[16 \zeta_3^2
+\frac{125 \pi ^2 \zeta_3}{3}-470 \zeta_3
+664 \zeta_5
+\frac{37729 \pi^6}{136080}-\frac{413 \pi ^4}{180}
   -\frac{6451 \pi ^2}{144}-\frac{2539}{12}
\nonumber\\& \hspace*{7mm}   
+t \left(-\frac{4}{3} \pi ^2 \zeta_3-214
   \zeta_3-240 \zeta_5+\frac{23 \pi ^4}{40}
   +\frac{119 \pi ^2}{4}+\frac{785}{8}\right)  
   +t^2 \left(102
   \zeta_3+\frac{83 \pi ^4}{360}-\frac{35 \pi ^2}{4}-\frac{507}{8}\right) 
\nonumber\\& \hspace*{7mm}   
+t^3 \left(-24 \zeta_3+\frac{3 \pi ^2}{2}+27\right) 
+t^4 \left(\frac{\pi ^2}{12}-\frac{17}{2}\right) +\frac{3t^5}{2}  -\frac{t^6}{6} \Biggr] 
\nonumber\\& \hspace*{0mm}
+ C_F^2 n_f \Biggl[ 
  \frac{41077}{972} - \frac{416}{9} \zeta_5+ \frac{13184}{81} \zeta_3
    -\frac{31729 \pi^2}{1944} - \frac{19\pi^2}{27} \zeta_3
     -\frac{331 \pi^4}{972}
\nonumber\\& \hspace*{13mm}  
   + t \bigg( \frac{3121}{108}-\frac{610}{9} \zeta_3
           +\frac{809\pi^2}{81} + \frac{7\pi^4}{45} \bigg)
\nonumber\\& \hspace*{13mm}  
   + t^2 \bigg(-\frac{12815}{324} + \frac{70}{9} \zeta_3 - \frac{56\pi^2}{27} \bigg)
   + t^3 \bigg(\frac{410}{27} + \frac{5\pi^2}{27} \bigg)
   - \frac{25}{9}t^4
   + \frac{2}{9}t^5
 \Biggr]
\nonumber\\& \hspace*{0mm}   
+ C_A C_F^2 \Biggl[ \frac{296 \zeta_3^2}{3}
-\frac{3751 \pi ^2 \zeta_3}{54}-\frac{18770 \zeta_3}{27}
-\frac{2756 \zeta_5}{9} 
-\frac{3169 \pi ^6}{17010}
-\frac{4943 \pi^4}{9720}+\frac{538835 \pi ^2}{3888}
+\frac{415025}{648}
\nonumber\\& \hspace*{15mm} 
+t \left(-\frac{5}{3} \pi ^2 \zeta_3+\frac{2441 \zeta_3}{3}
+120 \zeta_5+\frac{9 \pi ^4}{10}-\frac{5630 \pi ^2}{81}
-\frac{13805}{24}\right) 
\nonumber\\& \hspace*{15mm}  
+t^2 \left(-\frac{1807 \zeta_3}{9}-\frac{17 \pi ^4}{90}
+\frac{251 \pi^2}{27}+\frac{206317}{648}\right)  
+t^3 \left(26 \zeta_3
-\frac{\pi^2}{54}-\frac{2585}{27}\right) 
+t^4 \left(\frac{299}{18}
-\frac{\pi ^2}{3}\right) 
-\frac{11 t^5}{9}
\Biggr]
\nonumber\\& \hspace*{0mm}  
+ C_A^2 C_F \Biggl[-\frac{1136 \zeta_3^2}{9}
+\frac{208 \pi ^2 \zeta_3}{9}
+\frac{505087 \zeta_3}{486}
-\frac{434 \zeta_5}{9} -\frac{769 \pi^6}{5103}+\frac{22157 \pi ^4}{9720}
-\frac{412315 \pi ^2}{4374}-\frac{51082685}{52488}
\nonumber\\& \hspace*{15mm}
+t \left(\frac{44 \pi ^2 \zeta_3}{9}
-\frac{17464 \zeta_3}{27}+136 \zeta_5
-\frac{47 \pi ^4}{54}+\frac{8683 \pi ^2}{243}
+\frac{1045955}{1458}\right) 
\nonumber\\& \hspace*{15mm}
+ t^2 \left(88 \zeta_3
-\frac{11 \pi^4}{45}+\frac{13 \pi ^2}{27}-\frac{18682}{81}\right) 
+t^3 \left(\frac{2869}{81}
-\frac{22 \pi ^2}{27}\right)-\frac{121 t^4}{54}
\Biggr]
\nonumber\\& \hspace*{0mm}
+ C_A C_F n_f \Biggl[ \frac{2 \pi ^2 \zeta_3}{9}-\frac{4288 \zeta_3}{27}
-\frac{4 \zeta_5}{3} +\frac{\pi ^4}{486}+\frac{115555 \pi ^2}{4374}
+\frac{1700171}{6561}
\nonumber\\& \hspace*{19mm}
+\left(\frac{724 \zeta_3}{9}
+\frac{11 \pi ^4}{135}-\frac{2932 \pi^2}{243}-\frac{154919}{729}\right) t
\nonumber\\& \hspace*{19mm}
+\left(-8 \zeta_3+\frac{8 \pi^2}{9}+\frac{5876}{81}\right) t^2
+\left(\frac{4 \pi ^2}{27}-\frac{974}{81}\right) t^3
+\frac{22 t^4}{27} \Biggr]
\nonumber\\& \hspace*{0mm}
+C_F n_f^2 \Biggl[-\frac{416 \zeta_3}{243} -\frac{47 \pi^4}{1215}
-\frac{412 \pi ^2}{243}-\frac{190931}{13122} 
+t \left(\frac{16 \zeta_3}{27}+\frac{76 \pi ^2}{81}+\frac{9838}{729}\right) 
+t^2 \left(-\frac{406}{81}
-\frac{4 \pi^2}{27}\right) 
\nonumber\\& \hspace*{20mm}
+\frac{76 t^3}{81}
-\frac{2 t^4}{27}
 \Biggr]
\nonumber\\& \hspace*{0mm}
 + \frac{ C_F \left(N^2-4\right) N_{j,v}}{N} \Biggl[ \frac{14 \zeta_3}{3}
 -\frac{80 \zeta_5}{3}-\frac{\pi ^4}{90}+\frac{5 \pi ^2}{3}+4 \Biggr]
\Biggr\} \qquad + O(a_s^4),
\end{align}
where $t = \ln(Q_E^2/\mu^2) = \ln((-Q^2-i\epsilon)/\mu^2)= \ln(Q^2/\mu^2)-i\pi$.
The quantity $N_{j,v}$ is defined as 
\begin{equation}
\label{eq:N.f.v}
N_{j,v} \equiv \frac{\sum_q e_q}{e_j} \, .
\end{equation}
and is needed for graphs first encountered at $a_s^3$ where the quark
line at the electromagnetic current is in an internal loop instead of
being connected to the external lines.
Our own calculations are for the coefficient to order $a_s^2$,
  and are in agreement with the results first obtained from the
  same starting point by \citet{Idilbi:2006dg}.
Unlike $\gamma_K$ and $\gamma_j$, we cannot extract the result for
$H^{\text{Sud}}$ at order $a_s^3$ using the calculations in Ref.\
\cite{Moch:2005id} because the three-loop form factor given there
includes the pole terms but not the constant term as a function of $\epsilon$. To 
get the order $a_s^3$ contribution, the above steps may be straightforwardly 
repeated using the full three-loop result in Eq.~(5.4) of 
Ref.~\cite{Gehrmann:2010ue}.
Their results for $H^{\text{Sud}}$ are given in their Eqs.\
  (7.4), (7.5), and (7.8).  To make our Eq.\ (\ref{eq:Hsud}) a
  complete reference for the current state of knowledge, we have
  copied their order $a_s^3$ term.
\end{widetext}

In their paper reporting the form-factor calculations that we use,
\citet{Moch:2005id} also obtain the anomalous dimension, for the
quantity they call $A$, which equals $\gamma_K/2$.  Although they use the
same notation $A$ as in the CSS1 formulation, they have not performed
the CSS1 redefinition of the factors, and so their $A$ matches our
$\gamma_K$.  There appears to be no quantity calculated in
\cite{Moch:2005id} that corresponds directly to $\gamma_j$.

\section{RG evolution of TMD parton densities and TMD fragmentation functions}
\label{sec:PDFvFF}

The RG evolution of the TMD parton densities and TMD fragmentation functions is 
determined by their ultra-violet renormalization factors.  In turn, the 
renormalization factors are determined completely from the virtual graphs at the
vertices for the operators defining the TMD functions.  These are exactly the same 
graphs as for the square of the absolute value of any collinear factor in the form 
factor case. Therefore, the renormalization factor for a TMD function is the same 
as the square of the absolute value of the renormalization factor of the 
corresponding collinear factor for the form factor.  By the results of App.\
\ref{sec:TP}, this square has the same value independently of whether the Wilson 
lines are future- or past-pointing and of whether the quark is initial-state or final-
state. Thus the anomalous dimensions for the TMD functions are the same for 
TMD fragmentation functions and TMD parton densities, and they are also the 
same for the unpolarized TMD parton densities for SIDIS, with their past-pointing 
Wilson lines, and for the TMD parton densities for DY, with their future-pointing 
Wilson lines.

Hence in the RG equation \eqref{eq:RG.TMD.pdf} obeyed by the
  TMD parton densities, the anomalous dimensions $\gamma_j$ and $\gamma_K$ are
  the same as in the RG equation \eqref{eq:evo.C} for the collinear
  factors for the quark form factor.  Similarly the same anomalous
  dimensions are used in the RG equation for all the TMD fragmentation
  functions.

These relations have been known for some time from low-order
calculations, but the present paper is the first place we
know of where they are explicitly shown to be true generally. It is an
especially important result because it means the complete evolution
factor on the next-to-last line of Eq.~\eqref{eq:soln.2} is strongly
universal.

Note that these results do \emph{not} imply equality for the
coefficients $C^{\rm PDF}$ and $C^{\rm FF}$ that relate TMD functions
and the corresponding collinear functions; superscripts ``PDF'' and
``FF'' should be kept there.

\section{Values of Drell-Yan and SIDIS quantities}
\label{sec:DY.values}

In this section, we show in detail how to obtain values of
the coefficients at order $a_s^2$ for the Drell-Yan process
starting from results for collinear factorization and for the
  quark form factor.

\subsection{Hard factor}

Since the graphs and subtractions are the same, the hard factor for
Drell-Yan scattering is obtained from the square of the hard factor
for the time-like factor:
\begin{widetext}
\begin{equation}
  H_{j \bar{\jmath}}^{\rm DY}(Q,\mu;a_s(\mu)) 
  = e_j^2 \left| H_{j}^{\text{Sud, TL}}(Q^2)\right|^2 
  = e_j^2 \left| H_{j}^{\text{Sud}}(-Q^2-i\epsilon)\right|^2 
\, .
\end{equation}
From Eq.\ (\ref{eq:Hsud}) we find
\begin{align}
  \label{eq:HDY}
  \frac{1}{e_j^2} H_{j \bar{\jmath}}^{\rm DY}(Q,\mu;a_s(\mu)) ={}&   
      1
      + C_F a_s 
        \left( - 16 + \frac{7\pi^2}{3} + 6T - 2T^2 \right)
\nonumber\\& \hspace*{-30mm}
      + a_s^2 \Biggl\{
        C_F^2 \left[
                   \frac{511}{4}
                   - \frac{83\pi^2}{3}
                   + \frac{67\pi^4}{30}
                   - 60 \zeta_3
                   + T \left( -93 + 10\pi^2 + 48 \zeta_3 \right)
                   + T^2 \left( -\frac{14\pi^2}{3} + 50 \right) 
                   -12 T^3
                   + 2 T^4
        \right]
\nonumber\\ & \hspace*{-20mm}
         + C_F C_A  \left[
               - \frac{51157}{324}
               + \frac{1061 \pi^2}{54}
               - \frac{8 \pi^4}{45}
               + \frac{626}{9} \zeta_3 
               + T \left( \frac{2545}{27} - \frac{44 \pi^2}{6} - 52 \zeta_3 \right)
               + T^2 \left( \frac{2\pi^2}{3} - \frac{233}{9} \right)
               + \frac{22}{9} T^3 
        \right]
\nonumber\\&\hspace*{-20mm}
         +  n_f C_F \left[
                    \frac{4085}{162}
                    - \frac{91 \pi^2}{27}
                    + \frac{4}{9} \zeta_3
                    + T \left( \frac{8 \pi^2}{9} - \frac{418}{27} \right)
                    + \frac{38}{9} T^2 
                    - \frac{4}{9} T^3
        \right]
      \Biggr\}
\nonumber\\& \hspace*{-30mm}
 + a_s^3 
\Biggl\{C_F^3 \Biggl[
	32 \zeta_3^2-\frac{140 \pi^2 \zeta_3}{3}
	-460 \zeta_3+1328 \zeta_5+\frac{27403 \pi^6}{17010}
	-\frac{346 \pi^4}{15}+\frac{4339 \pi^2}{36}-\frac{5599}{6}  
\nonumber\\&\hspace*{-13mm}
	+ T \left( \frac{304 \pi^2 \zeta_3}{3}
	-992 \zeta_3-480 \zeta_5+\frac{109 \pi^4}{15}
	-89 \pi^2+\frac{1495}{2} \right)
	+ T^2 \left( 408 \zeta_3-\frac{67 \pi^4}{15}+\frac{220 \pi^2}{3}
   	           -\frac{1051}{2} \right)
\nonumber\\&\hspace*{-13mm}
	+ T^3 \left( -96 \zeta_3-20 \pi^2+222  \right)
	+ T^4 \left( \frac{14 \pi^2}{3}-68 \right)
	+ 12 T^5
	-\frac{4 T^6}{3}
	\Biggr]
\nonumber\\&\hspace*{-20mm}
+ C_A C_F^2 \Biggl[  
	\frac{592 \zeta_3^2}{3}+\frac{1690 \pi^2 \zeta_3}{9}	
	-\frac{52564 \zeta_3}{27}-\frac{5512 \zeta_5}{9}
	-\frac{1478 \pi^6}{1701}+\frac{92237 \pi^4}{2430}
	-\frac{406507 \pi^2}{972}+\frac{824281}{324}
\nonumber\\&\hspace*{-5mm}
	+ T \left( -116 \pi^2 \zeta_3+2252 \zeta_3
	+240 \zeta_5-\frac{1694 \pi^4}{135}
	+\frac{24268 \pi^2}{81}-\frac{14269}{6}  \right)
\nonumber\\&\hspace*{-5mm}		
	+ T^2 \left( -\frac{5644 \zeta_3}{9}+\frac{86 \pi^4}{45}
	-\frac{3376 \pi^2}{27}+\frac{208099}{162}  \right)
\nonumber\\&\hspace*{-5mm}		
	+ T^3 \left( 104 \zeta_3+\frac{526 \pi^2}{27}-\frac{10340}{27} \right)
	+ T^4 \left( \frac{598}{9}-\frac{4 \pi^2}{3} \right)
	-\frac{44 T^5}{9} 
	\Biggr]
\nonumber\\&\hspace*{-20mm}
+ C_A^2 C_F \Biggl[ 
	-\frac{2272 \zeta_3^2}{9}-\frac{1168 \pi ^2 \zeta_3}{9}
	+\frac{505087 \zeta_3}{243}-\frac{868 \zeta_5}{9}
	+\frac{4784 \pi^6}{25515}-\frac{4303 \pi^4}{4860}
	+\frac{596513 \pi^2}{2187}-\frac{51082685}{26244}
\nonumber\\&\hspace*{-5mm}	
	+ T \left( \frac{88 \pi^2 \zeta_3}{9}-\frac{34928 \zeta_3}{27}
	+272 \zeta_5+\frac{85 \pi^4}{27}-\frac{34276 \pi^2}{243}
	+\frac{1045955}{729} \right)
\nonumber\\&\hspace*{-5mm}		
	+ T^2 \left( 176 \zeta_3-\frac{22 \pi^4}{45}
	+\frac{752 \pi^2}{27}-\frac{37364}{81} \right)
	+ T^3 \left( \frac{5738}{81}-\frac{44 \pi^2}{27} \right)
	-\frac{121 T^4}{27}
	\Biggr]
\nonumber\\&\hspace*{-20mm}
C_A C_F n_f \Biggl[   
	\frac{148 \pi ^2 \zeta_3}{9}-\frac{8576 \zeta_3}{27}
	-\frac{8 \zeta_5}{3}-\frac{35 \pi^4}{243}
	-\frac{201749 \pi^2}{2187}+\frac{3400342}{6561}
\nonumber\\&\hspace*{-5mm}	
	+ T \left( \frac{1448 \zeta_3}{9}-\frac{98 \pi^4}{135}
	+\frac{11668 \pi^2}{243}-\frac{309838}{729} \right)
\nonumber\\&\hspace*{-5mm}		
	+ T^2 \left( -16 \zeta_3-8 \pi ^2+\frac{11752}{81} \right)
	+ T^3 \left( \frac{8 \pi^2}{27}-\frac{1948}{81} \right)
	+ \frac{44 T^4}{27}
	\Biggr]
\nonumber\\&\hspace*{-20mm}
+C_F n_f^2 \Biggl[  
	-\frac{832 \zeta_3}{243}+\frac{86 \pi^4}{1215}
	+\frac{1612 \pi^2}{243}-\frac{190931}{6561}
	+ T \left( \frac{32 \zeta_3}{27}-\frac{304 \pi^2}{81}
	+\frac{19676}{729}  \right)
\nonumber\\&\hspace*{-5mm}
	+ T^2 \left( \frac{16 \pi^2}{27}-\frac{812}{81} \right)
	+ \frac{152 T^3}{81}
	-\frac{4 T^4}{27}
	\Biggr]
\nonumber\\&\hspace*{-20mm}
+C_F^2 n_f \Biggl[  
	-\frac{148}{9} \pi^2 \zeta_3+\frac{26080 \zeta_3}{81}
	-\frac{832 \zeta_5}{9}-\frac{1463 \pi^4}{243}
	+\frac{13705 \pi^2}{243}-\frac{56963}{486}
\nonumber\\&\hspace*{-5mm}
	+ T \left( -\frac{1208 \zeta_3}{9}+\frac{332 \pi^4}{135}
	-\frac{4060 \pi^2}{81}+\frac{6947}{27} \right)
	+ T^2 \left( \frac{136 \zeta_3}{9}+\frac{520 \pi^2}{27}
	-\frac{14948}{81} \right)
\nonumber\\&\hspace*{-5mm}
	+ T^3 \left( \frac{1676}{27}-\frac{76 \pi^2}{27} \right)
	-\frac{100 T^4}{9}
	+ \frac{8 T^5}{9}
	\Biggr]
\nonumber\\&\hspace*{-20mm}
+C_F N_{j,v} \Biggl[  
	\frac{28 \zeta_3 N}{3}-\frac{160 \zeta_5 N}{3}
	-\frac{112 \zeta_3}{3 N}+\frac{640 \zeta_5}{3 N}
	-\frac{\pi^4 N}{45}+\frac{10 \pi ^2 N}{3}
	+8 N+\frac{4 \pi^4}{45 N}-\frac{40 \pi^2}{3 N}-\frac{32}{N}
	\Biggr] \Biggr\} ~ + O(a_s^4) \, .
\end{align}
For SIDIS, we get
\begin{align}
  \label{eq:HSIDIS}
  \frac{1}{e_j^2} H_{j \bar{\jmath}}^{\rm SIDIS}(Q,\mu;a_s(\mu)) ={}&   
      1
      + C_F a_s 
          \left(-16+\frac{\pi ^2}{3}+6 T-2 T^2\right) 
\nonumber\\& \hspace*{-30mm}
+a_s^2 
\Biggl\{C_F^2 \left[\frac{511}{4}+\frac{13 \pi ^2}{3} 
-\frac{13 \pi ^4}{30}-60 \zeta_3
+ T (-93 -2 \pi ^2+48 \zeta_3) + T^2 \Biggl(-\frac{2 \pi ^2}{3} + 50 \Biggr)
-12 T^3+2 T^4\right]  +
\nonumber\\ & \hspace*{-20mm}
+C_A C_F \left[-\frac{51157}{324}-\frac{337 \pi ^2}{54}+\frac{22 \pi ^4}{45}
+\frac{626 \zeta_3}{9}
+ T \Biggl( \frac{2545}{27}+\frac{22 \pi ^2 }{9} -52 \zeta_3\Biggr) 
+ T^2 \left(\frac{2 \pi ^2}{3} -\frac{233}{9} \right) +
\frac{22 T^3}{9} \right]
\nonumber\\&\hspace*{-20mm}
+ C_F n_f \left[ \frac{4085}{162}+\frac{23 \pi ^2}{27} 
+\frac{4 \zeta_3}{9} - T \left( \frac{418}{27}
+ \frac{4 \pi ^2}{9} \right)+\frac{38 T^2}{9}
-\frac{4 T^3}{9} \right] \Biggr\}
\nonumber\\&  \hspace*{-30mm}
 + a_s^3 
\Biggl\{C_F^3 \Biggl[  
	32 \zeta_3^2+\frac{220 \pi^2 \zeta_3}{3}
	-460 \zeta_3+1328 \zeta_5+\frac{1625 \pi^6}{3402}
	+\frac{4 \pi^4}{15}-\frac{4859 \pi^2}{36}-\frac{5599}{6}
\nonumber\\&\hspace*{-13mm}
	+ T \left( \frac{16 \pi^2 \zeta_3}{3}-992 \zeta_3
	-480 \zeta_5-\frac{11 \pi^4}{15}+97 \pi^2+\frac{1495}{2} \right)
	+ T^2 \left( 408 \zeta_3+\frac{13 \pi^4}{15}
	            - \frac{80 \pi^2}{3}-\frac{1051}{2} \right)
\nonumber\\&\hspace*{-13mm}
	+ T^3 \left( -96 \zeta_3+4 \pi^2+222 \right)
	+ T^4 \left( \frac{2 \pi^2}{3}-68  \right)
	+ 12 T^5
	-\frac{4 T^6}{3}
		\Biggr]
\nonumber\\&\hspace*{-20mm}
+C_A C_F^2 \Biggl[ 
	\frac{592 \zeta_3^2}{3}-\frac{382 \pi^2 \zeta_3}{3}
	-\frac{52564 \zeta_3}{27}-\frac{5512 \zeta_5}{9}
	-\frac{2476 \pi^6}{8505}-\frac{14503 \pi^4}{2430}
	+\frac{292367 \pi^2}{972}
	+\frac{824281}{324}
\nonumber\\&\hspace*{-5mm}
	+ T \left( -12 \pi^2 \zeta_3+2252 \zeta_3
	+240 \zeta_5+\frac{496 \pi^4}{135}-\frac{13088 \pi^2}{81}
	-\frac{14269}{6} \right)
\nonumber\\&\hspace*{-5mm}
	+ T^2 \left( -\frac{5644 \zeta_3}{9}-\frac{34 \pi^4}{45}
	+\frac{608 \pi^2}{27}+ \frac{208099}{162}  \right)
	+ T^3 \left( 104 \zeta_3-\frac{2 \pi^2}{27}-\frac{10340}{27} \right)
\nonumber\\&\hspace*{-5mm}
	+ T^4 \left( \frac{598}{9}-\frac{4 \pi^2}{3} \right)
	-\frac{44 T^5}{9}
	\Biggr]
\nonumber\\&\hspace*{-20mm}
+ C_A^2 C_F \Biggl[  
	-\frac{2272 \zeta_3^2}{9}+\frac{416 \pi^2 \zeta_3}{9}
	+\frac{505087 \zeta_3}{243}-\frac{868 \zeta_5}{9}
	-\frac{1538 \pi^6}{5103}+\frac{22157 \pi^4}{4860}
	-\frac{412315 \pi^2}{2187}-\frac{51082685}{26244}
\nonumber\\&\hspace*{-5mm}
	+ T \left( \frac{88 \pi^2 \zeta_3}{9}-\frac{34928 \zeta_3}{27}
	+272 \zeta_5-\frac{47 \pi^4}{27}+\frac{17366 \pi^2}{243}
	+\frac{1045955}{729} \right)
\nonumber\\&\hspace*{-5mm}
	+ T^2 \left( 176 \zeta_3-\frac{22 \pi^4}{45}
	+\frac{26 \pi^2}{27}-\frac{37364}{81}  \right)
	+ T^3 \left( \frac{5738}{81}-\frac{44 \pi^2}{27} \right)
	-\frac{121 T^4}{27}
		\Biggr]
\nonumber\\&\hspace*{-20mm}
C_A C_F n_f \Biggl[  
	\frac{4 \pi^2 \zeta_3}{9}-\frac{8576 \zeta_3}{27}
	-\frac{8 \zeta_5}{3}+\frac{\pi^4}{243}
	+\frac{115555 \pi^2}{2187}+\frac{3400342}{6561}
\nonumber\\&\hspace*{-5mm}
	+ T \left( \frac{1448 \zeta_3}{9}+\frac{22 \pi^4}{135}
	-\frac{5864 \pi^2}{243}-\frac{309838}{729}  \right)
	+ T^2 \left( -16 \zeta_3+\frac{16 \pi ^2}{9}+\frac{11752}{81} \right)
\nonumber\\&\hspace*{-5mm}
	+ T^3 \left( \frac{8 \pi^2}{27}-\frac{1948}{81} \right)
	+\frac{44 T^4}{27}
	\Biggr]
\nonumber\\&\hspace*{-20mm}
+C_F n_f^2 \Biggl[
	-\frac{832 \zeta_3}{243}-\frac{94 \pi^4}{1215}
	-\frac{824 \pi^2}{243}-\frac{190931}{6561}  
	+ T \left( \frac{32 \zeta_3}{27}+\frac{152 \pi^2}{81}
	+\frac{19676}{729} \right)
\nonumber\\&\hspace*{-5mm}
	+ T^2 \left( -\frac{812}{81}-\frac{8 \pi^2}{27} \right)
	+ \frac{152 T^3}{81} 
	-\frac{4 T^4}{27}
	\Biggr]
\nonumber\\&\hspace*{-20mm}
+C_F^2 n_f \Biggl[  
	-\frac{4}{3} \pi^2 \zeta_3+\frac{26080 \zeta_3}{81}
	-\frac{832 \zeta_5}{9}-\frac{131 \pi^4}{243}-\frac{8567 \pi^2}{243}
	-\frac{56963}{486}
\nonumber\\&\hspace*{-5mm}
	+ T \left(-\frac{1208 \zeta_3}{9}+\frac{32 \pi^4}{135}
	+\frac{1904 \pi^2}{81}+\frac{6947}{27}\right)
\nonumber\\&\hspace*{-5mm}
	+ T^2 \left( \frac{136 \zeta_3}{9}-\frac{152 \pi^2}{27}
	-\frac{14948}{81} \right)
	+ T^3 \left( \frac{1676}{27}+\frac{20 \pi^2}{27} \right)
	-\frac{100 T^4}{9} 
	+ \frac{8 T^5}{9}
	\Biggr]
\nonumber\\&\hspace*{-20mm}
+C_F N_{j,v} \Biggl[  
	\frac{28 \zeta_3 N}{3}-\frac{160 \zeta_5 N}{3}
	-\frac{112 \zeta_3}{3 N}+\frac{640 \zeta_5}{3 N}
	-\frac{\pi^4 N}{45}
	+\frac{10 \pi^2 N}{3}+8 N+\frac{4 \pi^4}{45 N}
	-\frac{40 \pi^2}{3 N}-\frac{32}{N}
	\Biggr] 
\Biggr\} + O(a_s^4) \, .
\end{align}
In both of these equations,
$T = \ln (Q^2/\mu^2)$, and $N_{j,v}$ is defined by Eq.\ (\ref{eq:N.f.v}).
With $n_f =3$, the ratio of the Drell-Yan to SIDIS hard factors is
\begin{equation}
\frac{H_{j \bar{\jmath}}^{\rm DY}}{H_{j \bar{\jmath}}^{\rm SIDIS}}
= 1 + 2.0944 \, \alpha_s(\mu) + 5.96498 \, \alpha_s(\mu)^2 + 
18.6104 \, \alpha_s (\mu)^3 + O(\alpha_s^4) \, ,
\end{equation}
and we have verified that we match Eq.~(4.4) of 
Ref.~\cite{Moch:2005id} for $n_f = 4$.

In our later calculations, we will need the coefficients of the Drell-Yan hard
factor at $T=0$, i.e., with $\mu=Q$ or $C_2=1$.  So we write 
\begin{equation}
  \frac{1}{e_j^2} H_{j \bar{\jmath}}^{\rm DY}(Q,Q;a_s(Q))
  = 1 + \sum_{n=1}^{\infty} a_s^n \hat{H}_{j \bar{\jmath}}^{\text{DY }(n)} ,
\end{equation}
and we have
\begin{subequations}
\begin{align}
  \hat{H}_{j \bar{\jmath}}^{\text{DY }(1)}
  ={}&
    C_F \left( - 16 + \frac{7\pi^2}{3}  \right),
\label{eq:H.hat.1}
\\
  \hat{H}_{j \bar{\jmath}}^{\text{DY }(2)}
  ={}&
        C_F^2 \left[
                   \frac{511}{4}
                   - \frac{83\pi^2}{3}
                   + \frac{67\pi^4}{30}
                   - 60 \zeta_3
        \right]
         + C_F C_A  \left[
               - \frac{51157}{324}
               + \frac{1061 \pi^2}{54}
               - \frac{8 \pi^4}{45}
               + \frac{626}{9} \zeta_3 
        \right]
\nonumber\\&
         +  n_f C_F \left[
                    \frac{4085}{162}
                    - \frac{91 \pi^2}{27}
                    + \frac{4}{9} \zeta_3 
        \right] ,
\label{eq:H.hat.2}
\\
  \hat{H}_{j \bar{\jmath}}^{\text{DY }(3)}
  ={}&
  C_F^3 \Biggl[  
	32 \zeta_3^2+\frac{220 \pi^2 \zeta_3}{3}
	-460 \zeta_3+1328 \zeta_5+\frac{1625 \pi^6}{3402}
	+\frac{4 \pi^4}{15}-\frac{4859 \pi^2}{36}-\frac{5599}{6}
      \Biggr]
\nonumber\\&
+C_A C_F^2 \Biggl[ 
	\frac{592 \zeta_3^2}{3}-\frac{382 \pi^2 \zeta_3}{3}
	-\frac{52564 \zeta_3}{27}-\frac{5512 \zeta_5}{9}
	-\frac{2476 \pi^6}{8505}-\frac{14503 \pi^4}{2430}
\nonumber\\&\hspace*{15mm}
	+\frac{292367 \pi^2}{972}
	+\frac{824281}{324}
    \Biggr]
\nonumber\\&
+C_A^2 C_F \Biggl[  
	-\frac{2272 \zeta_3^2}{9}+\frac{416 \pi^2 \zeta_3}{9}
	+\frac{505087 \zeta_3}{243}-\frac{868 \zeta_5}{9}
	-\frac{1538 \pi^6}{5103}+\frac{22157 \pi^4}{4860}
\nonumber\\&\hspace*{15mm}	
	-\frac{412315 \pi^2}{2187}-\frac{51082685}{26244}
    \Biggr]
\nonumber\\&
C_A C_F n_f \Biggl[  
	\frac{4 \pi^2 \zeta_3}{9}-\frac{8576 \zeta_3}{27}
	-\frac{8 \zeta_5}{3}+\frac{\pi^4}{243}
	+\frac{115555 \pi^2}{2187}+\frac{3400342}{6561}
	\Biggr]
\nonumber\\&
+C_F n_f^2 \Biggl[
	-\frac{832 \zeta_3}{243}-\frac{94 \pi^4}{1215}
	-\frac{824 \pi^2}{243}-\frac{190931}{6561}  
	\Biggr]
\nonumber\\&
+C_F^2 n_f \Biggl[  
	-\frac{4}{3} \pi^2 \zeta_3+\frac{26080 \zeta_3}{81}
	-\frac{832 \zeta_5}{9}-\frac{131 \pi^4}{243}-\frac{8567 \pi^2}{243}
	-\frac{56963}{486}
	\Biggr]
\nonumber\\&
+C_F N_{j,v} \Biggl[  
	\frac{28 \zeta_3 N}{3}-\frac{160 \zeta_5 N}{3}
	-\frac{112 \zeta_3}{3 N}+\frac{640 \zeta_5}{3 N}
	-\frac{\pi^4 N}{45}
\nonumber\\&\hspace*{15mm}		
	+\frac{10 \pi^2 N}{3}+8 N+\frac{4 \pi^4}{45 N}
	-\frac{40 \pi^2}{3 N}-\frac{32}{N}
	\Biggr] .
\label{eq:H.hat.3}
\end{align}
\label{eq:H.hat}
\end{subequations}

\subsection{RG coefficients}

Values for $\gamma_j$ and $_K$ equal those for the quark Sudakov form
factor, given our choice of normalizations, and were already given in
Eqs.\ \eqref{eq:gammaj} and \eqref{eq:gammaK}.

\subsection{CSS evolution coefficient}

Values for $\tilde{K}\left(\Tsc{b} ; \mu \right)$ are obtained from
Eqs.~\eqref{eq:B.CSS1}, \eqref{eq:gammaK}, and \eqref{eq:HDY},
 and the renormalization group relation
\begin{equation}
\tilde{K}(\bstarsc;\muQ) = \tilde{K}(\bstarsc;\mubstar)  - \int_{\mubstar}^{\muQ}  \frac{ \diff{\mu'} }{ \mu' } \gamma_K(a_s(\mu')) \, .
\end{equation}
To use this equation to obtain terms in the perturbative
expansion of $\tilde{K}$, the coupling $a_s(\mu')$ must be expanded
in powers of $a_s(\mu_Q)$.
We utilize the results up to order $a_s^2$ for $B_{\text{CSS1, DY}}(a_s;2e^{-\gammae},1)$ from Ref.~\cite{Davies:1984hs}, and obtain
\begin{align}
\label{eq:K}
\tilde{K}(\Tsc{b}; \mu)   
={}& - 8 C_F a_s(\mu) \ln \left( \frac{b_T \mu}{2 e^{-\gammae}} \right)  \nonumber \\
  & +  8 C_F a_s(\mu)^2 \left[ \left( \frac{2}{3} n_f  - \frac{11}{3} C_A \right) \ln^2 \left( \frac{\Tsc{b} \mu}{2 e^{-\gammae}} \right) 
                        \right. 
\nonumber \\ 
 & \qquad \qquad \qquad
+ \left. 
             \left( -\frac{67}{9}C_A + \frac{\pi^2}{3} C_A +  \frac{10}{9} n_f \right) \ln \left( \frac{\Tsc{b} \mu}{2 e^{-\gammae}} \right) 
               + \left( \frac{7}{2} \zeta_3 - \frac{101}{27} \right)
               C_A 
               + \frac{14}{27}  n_f
      \right]  \, \nonumber \\
    &  + O(a_s^3) \, .
\end{align}
By differentiating with respect to $\Tsc{b}$, one may easily
verify that this is consistent with the so-far unused relation Eq.\
(\ref{eq:A.CSS1}), and the value of $A_{\rm CSS1}(a_s; 2e^{-\gammae})$ in
Ref.~\cite{Davies:1984hs}.

The value of $\tilde{K}$ up to order $a_s^3$ is given by
calculations of the soft factor reported by
\citet{Li:2016ctv}.\footnote{This
  result was independently calculated and confirmed by
  \citet{Vladimirov:2016dll} by a use of a conformal transformation on a
  Wilson line matrix element, to relate its rapidity divergence to a UV
  divergence; by the use of a correspondence of rapidity renormalization
  between soft factors and TMD functions \cite{Echevarria:2016scs}, there
  is obtained a result for (an equivalent of) $\tilde{K}$ from a known UV
  anomalous dimension.}
The correspondence with the CSS2 version of factorization is quite
non-trivial.  This is because of a different organization of factors
and a different approach to rapidity divergences, in the form given
by \citet{Li:2016axz}.  We obtain the correspondence in App.\
\ref{app:corr.Li.et.al}.  As shown there, $\tilde{K}$ equals the
right-hand side of Eq.\ (4) of Ref.\ \cite{Li:2016ctv}, and equals
the $\gamma_R$ of \cite{Li:2016axz}.  Then the actual perturbative
coefficients when $\mu= 2 e^{-\gammae}/b_T$ are in Eq.\ (9) of Ref.\
\cite{Li:2016ctv}, with the $\mu$ dependence given in terms of $\gamma_K$
by our Eq.\ (\ref{eq:RG.K}).  See also Ref.\ \cite{Luebbert:2016itl,
  Echevarria:2016scs} for other calculations of a differently
normalized version of $\tilde{K}$ at order $a_s^2$, again starting
from the operator definitions of the TMD parton densities and soft
function, and in agreement with Eq.\ (\ref{eq:K}).  

\subsection{Wilson coefficients $\tilde{C}$ for TMD quark density}
\label{sec:DY.from.MVV}

The coefficient functions $\tilde{C}$ in the new formalism can now be found from those 
of the old by using Eq.~\eqref{eq:C.CSS1}, which gives
\begin{equation}
\label{eq:C.CSS1.rev}
\tilde{C}^{\rm PDF}_{j/k}\xleft( \frac{x}{\xi},\bstarsc;\mubstar^2, \mubstar, a_s(\mubstar) \right)
  = \frac{ \tilde{C}^{\text{CSS1, DY}}_{j/k}\xleft(
           \frac{x}{\xi},\bstarsc;
           \mubstar^2, \mubstar, C_2, a_s(\mubstar) \right)}
         {  \sqrt{ (1/e_j^2) H^{\rm DY}_{j\bar{\jmath}}(\mubstar/C_2,
             \mubstar, a_s(\mubstar))}}
    \exp\xleft[ \tilde{K}(\bstarsc;\mubstar) \ln C_2
      \right] \, . 
\end{equation}
To get results for $\tilde{C}$ up to order $a_s^2$ in the new formalism, 
we use the order $a_s^2$ results for $H^{\rm DY}$ and $\tilde{K}$ from Eqs.~\eqref{eq:HDY} and~\eqref{eq:K}.
(Note that if the standard choice of $C_2 = 1$ is used, the exponential factor becomes trivial.)
The CSS1 coefficient functions have been obtained to order $a_s^2$ by 
Catani, Cieri, de Florian, Ferrera, and Grazzini (CCFFG) in Ref.~\cite{Catani:2012qa}.
The expansion coefficients for $\tilde{C}$ (and similarly for
$\tilde{C}^{\rm CSS1}$) are given in our usual notation:
\begin{equation}
\tilde{C}^{\rm PDF}_{a/b}(x,\bstarsc;\zeta,\mu,a_s(\mu))  
=  \delta_{ab} \delta(1 - x) +
\sum_{n=1}^\infty a_s(\mu)^n \tilde{C}^{{\rm PDF}, (n)}_{ab}
(x,\bstarsc;\zeta,\mu) \,,
\end{equation}
where we have restored general values of the arguments.

CCFFG express their results in terms of a function $\mathcal{H}^{\rm
  DY}_{f_1 f_2 \gets f_3 f_4}$, where $f_3$ and $f_4$ are the flavors
of partons in the collinear parton densities and $f_1$ and $f_2$ are the flavors
of partons that enter the hard scattering.  They make the specific
choice that $\mu=\sqrt{\zeta} = b_0/\bstarsc$, with
$b_0=2e^{-\gammae}$, i.e., $C_1=b_0$, $C_2=1$.  The $C$ coefficient 
functions in Ref.~\cite{Catani:2012qa} are expressed in terms of
$\mathcal{H}^{\rm DY}$ and a scheme-dependent function called
$H$ (not to be confused with the $H$ used in the present paper).  

The CCFFG $H$ is the vertex factor in
Ref.~\cite[Eq.~(7)]{Catani:2012qa}.  However, the $\tilde{C}$
functions in that  
formula are not necessarily connected to specific correlation functions for 
TMD functions, and so there remains a choice as to how perturbative 
parts are to be partitioned between different factors. One must choose a 
resummation scheme. By comparing with Eqs.~\eqref{eq:CSS.A.B},
\eqref{eq:fact.CSS2}, and \eqref{eq:ABequality} of this paper, it is clear that the 
CCFFG $\tilde{C}$ functions correspond to CSS1 $\tilde{C}$ functions if 
all non-zeroth order contributions to the CCFFG $H$ function are set to zero, 
while they are the CSS2 $\tilde{C}$ functions if $H$ is set equal to the
$H^{\text{DY}}$ functions of Eq.~\eqref{eq:fact.CSS2} and \eqref{eq:HDY}. (CCFFG define 
still another choice called the hard resummation scheme -- see the discussion of 
Eqs.(22-27) of Ref.~\cite{Catani:2012qa}.)

The reason CSS2 has a definite value for $H$ but CCFFG do not is 
that CSS2 uses a specific definition of the TMD functions; CCFFG only 
provide information that is determined from calculations relevant for
collinear factorization without reference to the definition of TMD
functions.

At order $a_s$, using Eqs.~(14)--(16) of
Ref.~\cite{Catani:2012qa} gives
\begin{subequations}
\begin{align}
  \tilde{C}^{\text{CSS1, DY, (1)}}_{q/q}(x,\bstarsc; b_0^2/\bstarsc^2,b_0/\bstarsc , C_2\mapsto1)
   ={}& 
   C_F \left[ \left( \pi^2 - 8 \right) \delta(1 - x)  + 2(1 - x) \right]  ,
\label{qqLO1}
\\
   \tilde{C}^{\text{CSS1, DY, (1)}}_{q/g}(x; b_0^2/\bstarsc^2, b_0/\bstarsc , C_2\mapsto1)
    ={}&
    2 x (1 - x) ,
\label{qqLO2}
\\
   \tilde{C}^{\text{CSS1, DY, (1)}}_{q/q'}(x)
   ={}&
   \tilde{C}^{\text{CSS1, DY, (1)}}_{q/\bar{q}}(x) 
   = \tilde{C}^{\text{CSS1, DY, (1)}}_{q/\bar{q}'}(x) = 0 \, ,
\label{qqLO3}
\end{align}
\end{subequations}
in agreement with the original results, Eqs.\ (3.25) and
  (3.26) of Ref.\ \cite{Collins:1984kg}.
Here, $q$ and $q'$ are quarks of different flavors. 
Note that in Ref.~\cite{Catani:2012qa}, the expansion parameter is
$\alpha_s/\pi$ rather than our $\alpha_s/(4\pi)$, so that the above
coefficients differ by a factor 4 from the corresponding coefficients
in Ref.~\cite{Catani:2012qa}.

For order-$a_s^2$, the same procedure gives, using Eqs.~(32), (34) and
(35) of Ref.~\cite{Catani:2012qa},
\begin{subequations}
\begin{align}
\tilde{C}^{\text{CSS1, DY, (2)}}_{q/q}(x; b_0^2/\bstarsc^2, b_0/\bstarsc , C_2\mapsto1)
={}& 
  8 {\cal H}^{\rm DY(2)}_{q\bar q\gets q\bar q}(x) 
   - 2 C_F^2
     \left[ \delta(1-x)  \f{(\pi^2-8)^2}{4}
            + \left(\pi^2-10\right)(1-x) - (1+x) \ln x
     \right] \;,
\label{qq22}
\displaybreak[0]\\
\tilde{C}^{\text{CSS1, DY, (2)}}_{q/g}(x; b_0^2/\bstarsc^2, b_0/\bstarsc , C_2\mapsto1)
={}&
  16 {\cal H}^{\rm DY(2)}_{q\bar q\gets qg}(x)
  - 2C_F
   \left[
      2x \ln x + 1-x^2 + \left( \pi^2-8 \right)x\,(1-x)
   \right] \;,
\label{qg22}
\displaybreak[0]\\
\tilde{C}^{\text{CSS1, DY, (2)}}_{q/\bar q}(x; b_0^2/\bstarsc^2, b_0/\bstarsc , C_2\mapsto1)
={}&
  16 {\cal H}^{\rm DY(2)}_{q\bar q\gets q q}(x) \;,
\label{qqp2}
\displaybreak[0]\\
\tilde{C}^{\text{CSS1, DY, (2)}}_{q/q'}(x; b_0^2/\bstarsc^2, b_0/\bstarsc , C_2\mapsto1)
={}&
  16 {\cal H}^{\rm DY(2)}_{q\bar q\gets q \bar{q}'}(x) \; ,
\displaybreak[0]\\
\tilde{C}^{\text{CSS1, DY, (2)}}_{q/\bar{q}'}(x; b_0^2/\bstarsc^2, b_0/\bstarsc , C_2\mapsto1)
={}&
  16 {\cal H}^{\rm DY(2)}_{q\bar q\gets q q'}(x)\; .
\label{qqp3}
\end{align}
\end{subequations}
where the formulas for the CCFFG $\mathcal{H}^{\rm DY(2)}$-functions are
given in Eqs.~(23)--(29) of Ref.~\cite{Catani:2012qa}.  

These expressions are given for the standard choice that the $\zeta$
and $\mu$ arguments of $\tilde{C}$ are set to $b_0^2/\bstarsc^2$,
$b_0/\bstarsc$.  Then from Eqs.\ \eqref{eq:H.hat} and \eqref{eq:C.CSS1.rev}, we 
find the CSS2 coefficients:
\begin{subequations}
\label{eq:C.CSS2}
\begin{align}
  \tilde{C}^{{\rm PDF}, (1)}_{q/q}(x,\bstarsc;b_0^2/\bstarsc^2,b_0/\bstarsc)
   ={}& 
   C_F \left[ - \frac{\pi^2}{6} \delta(1 - x)  + 2(1 - x) \right] ,
\label{qqLO1.CSS2}
\displaybreak[0]\\
   \tilde{C}^{{\rm PDF}, (1)}_{q/g}(x; b_0^2/\bstarsc^2, b_0/\bstarsc)
    ={}&
    2 x (1 - x) ,
\label{qqLO2.CSS2}
\displaybreak[0]\\
   \tilde{C}^{{\rm PDF}, (1)}_{q/q'}(x)
   ={}&
   \tilde{C}^{{\rm PDF}, (1)}_{q/\bar{q}}(x) 
   = \tilde{C}^{{\rm PDF}, (1)}_{q/\bar{q}'}(x) = 0 \, .
\label{qqLO3.CSS2}
%
\displaybreak[0]\\
%
\tilde{C}^{{\rm PDF}, (2)}_{q/q}(x; b_0^2/\bstarsc^2, b_0/\bstarsc)
={}& 
  8 {\cal H}^{\rm DY(2)}_{q\bar q\gets q\bar q}(x) 
   - 2 C_F^2
     \left[ \delta(1-x)  \f{(\pi^2-8)^2}{4}
            + \left(\pi^2-10\right)(1-x) - (1+x) \ln x
     \right] -
\nonumber \\
     & - C_F^2 \left(\frac{7\pi^2}{6} - 8\right)
             \left[ (\pi^2 - 8) \delta(1-x) + 2 (1-x) \right]
\nonumber \\
     & + \delta(1-x) \left[ - \frac{1}{2}\hat{H}_{j \bar{\jmath}}^{\text{DY }(2)}
                       + \frac{3}{8} 
                         \left( \hat{H}_{j \bar{\jmath}}^{\text{DY }(1)} \right)^2
                \right] ,
\label{qq22.CSS2}
\displaybreak[0]\\
\tilde{C}^{{\rm PDF}, (2)}_{q/g}(x; b_0^2/\bstarsc^2, b_0/\bstarsc)
={}&
  16 {\cal H}^{\rm DY(2)}_{q\bar q\gets qg}(x)
  - 2C_F
   \left[
      2x \ln x + 1-x^2 + \left( \frac{13 \pi^2}{6} -16 \right)x\,(1-x)
   \right] \;,
\label{qg22.CSS2}
\displaybreak[0]\\
\tilde{C}^{{\rm PDF}, (2)}_{q/\bar q}(x; b_0^2/\bstarsc^2, b_0/\bstarsc)
={}&
  16 {\cal H}^{\rm DY(2)}_{q\bar q\gets q q}(x) \;,
\label{qqp2.CSS2}
\displaybreak[0]\\
  \tilde{C}^{{\rm PDF}, (2)}_{q/q'}(x; b_0^2/\bstarsc^2, b_0/\bstarsc)
  ={}&
  16 {\cal H}^{\rm DY(2)}_{q\bar q\gets q \bar{q}'}(x) \; ,
\displaybreak[0]\\
  \tilde{C}^{{\rm PDF}, (2)}_{q/\bar{q}'}(x; b_0^2/\bstarsc^2, b_0/\bstarsc)
  ={}&
  16 {\cal H}^{\rm DY(2)}_{q\bar q\gets q q'}(x)\; .
\label{qqp3.CSS2}
\end{align}
\end{subequations}

To obtain results for the coefficients with general
values of $\zeta$ and $\mu$, which we do not present explicitly
  here,
one can use the evolution equations for
$\tilde{C}$.  These show that the dependence of $\tilde{C}$ in each
order of
$a_s$ is polynomial in $\ln\frac{\bstarsc\mu}{b_0}$ and
$\ln\frac{\bstarsc^2\zeta}{b_0^2}$, and the coefficients of the
logarithms can be deduced from the equations.  These equations are
\begin{align}
  \frac{ \diff{ \tilde{C}^{\rm PDF}_{a/b}(z,\bstarsc;\zeta,\mu,a_s(\mu)) } }
       { \diff{ \ln\sqrt{\zeta} } }
  ={}& 
  \tilde{K}(\bstarsc,\mu,a_s(\mu)) ~ \tilde{C}^{\rm PDF}_{a/b}(z,\bstarsc;\zeta,\mu,a_s(\mu)),
\\
  \frac{ \diff{ \tilde{C}^{\rm PDF}_{a/b}(z,\bstarsc;\zeta,\mu,a_s(\mu)) } }
       { \diff{ \ln\mu } }
  = {}&
  \left[ \gamma_j( a_s(\mu))
      - \frac12 \gamma_K(a_s(\mu)) \ln \frac{ \zeta }{ \mu^2 }
  \right]
  \tilde{C}^{\rm PDF}_{a/b}(z,\bstarsc;\zeta,\mu,a_s(\mu))
\nonumber\\ &
  - 2 \sum_k \int_z^1 \frac{ \diff{y} }{ y }
      \tilde{C}^{\rm PDF}_{a/k}(z/y,\bstarsc;\zeta,\mu,a_s(\mu))
      P_{kb}(y,a_s(\mu)) .
\end{align}
These can in turn be derived from the evolution equations \eqref{eq:CSS.evol}
and \eqref{eq:RG.TMD.pdf} for the TMD parton densities and the DGLAP
equation for the collinear parton densities,
\begin{equation}
\label{eq:DGLAP}
  \frac{ \diff{ f_{a/H}(z,\mu) } }
       { \diff{ \ln\mu } }
  =
  2 \sum_k \int_x^1 \frac{\diff{\xi} }{ \xi }
    \, P_{ak}(x/\xi, a_s(\mu))
    f_{k/H}(\xi;\mu).
\end{equation}

We have compared the values in Eqs.\ \eqref{eq:C.CSS2} with those
found in Ref.\ \cite{Echevarria:2016scs}, and found agreement.  
In making the comparison, the following points are important.  First
the identities
\begin{align}
  \text{Li}_2(z) + \text{Li}_2(1-z) ={}&
           - \ln(z) \ln(1-z) + \frac{\pi^2}{6},
\\
  \text{Li}_2(z) + \text{Li}_2(-z) ={}& \frac{1}{2} \text{Li}_2(z^2),
\\
  \text{Li}_2(z) + \text{Li}_2(1/z) ={}&
           - \frac{1}{2} \ln^2(-z) - \frac{\pi^2}{6},
\\
  \text{Li}_3(z) + \text{Li}_3(-z) ={}& \frac{1}{4} \text{Li}_3(z^2),
\\
  \text{Li}_3(z) + \text{Li}_3(1-z) + \text{Li}_3(1-1/z) ={}&
           \zeta_3 + \frac{1}{6} \ln^3 z 
           + \frac{\pi^2}{6} \ln z - \frac{1}{2} \ln^2z \ln(1-z),
\end{align}
are needed.
Here the polylogarithm functions are
\begin{eqnarray}
\text{Li}_2(z) & = & -\int_0^z \frac{\diff{} t}{t} \ln (1 - t) \, , \\
\text{Li}_3(z) & = & \int_0^1 \frac{\diff{} t}{t} \ln (t) \ln(1 - z t) \, .
\end{eqnarray}%
Second, our flavor-diagonal matching coefficient
$C^{\text{PDF}(2)}_{q/q}$, is the full matching coefficient.  The
apparently corresponding coefficient in Ref.\
\cite{Echevarria:2016scs} is $C^{(2,0)}_{q \gets q}$.  But in fact the
full matching coefficient is obtained by adding to this the term for
non-matching quark flavors $C^{(2,0)}_{q \gets q'}$.  A corresponding
remark applies to the $q \gets \bar{q}$ coefficient.
\end{widetext}

\section{Conclusions}
\label{sec:concl}

We conclude by summarizing and highlighting our main results. 

Firstly, we have established the mapping between quantities 
in the earlier CSS1 organization of factorization, for which
there are many previous calculations and fits, and the newer CSS2
method.
The results for CSS2 also apply to the SCET-based formalism of
\citet{GarciaEchevarria:2011rb}, since their TMD functions and
factorization formulae are equivalent to the CSS2 ones.  They also apply to the method of
\citet{Li:2016ctv, Li:2016axz}, provided that TMD functions are defined by
absorbing a square root of their soft factor into each beam
function, as we explain in App.\ \ref{app:corr.Li.et.al}.
Perturbative quantities in one formalism 
are directly related to those of the other with equations like (\ref{eq:A.CSS1}),
(\ref{eq:B.CSS1}), and (\ref{eq:C.CSS1.rev}). 
Furthermore, as regards the non-perturbative transverse-momentum dependence,
we have established that the $g$-functions like 
$g_K(\Tsc{b})$ and $g_{j/H}(x,\Tsc{b}{})$ are identical in CSS1 and in
CSS2. 
Therefore fits of these functions
obtained using CSS1 (e.g.,~\cite{Landry:2002ix, Konychev:2005iy})
may correctly be used in CSS2, and in the SCET formalisms of Refs.\
\cite{GarciaEchevarria:2011rb, Li:2016ctv, Li:2016axz}.

Secondly, we have shown in detail how to obtain the
perturbative quantities in the new formalisms from a combination of
calculations for $\Tsc{q}$ distributions in collinear factorization,
as in Refs.~\cite{Catani:2000vq,Catani:2012qa}, with calculations of
the dimensionally regulated massless quark form factor, as in Ref.\
\cite{Moch:2005id, Gehrmann:2010ue}.  It is quite non-trivial that
the anomalous dimensions $\gamma_j$ and $\gamma_K$ for TMD functions can be
obtained from the form factor alone.  We showed explicitly that the
results agree with those obtained directly from calculations
\cite{Echevarria:2016scs, Gehrmann:2012ze, Gehrmann:2014yya,
  Grozin:2014hna, Li:2016ctv, Grozin:2015kna, Luebbert:2016itl} of
the matrix elements of the operators, involving Wilson lines, that
are used in the definitions of the unsubtracted TMD functions and the soft
function.  Although some of our results appear to be known in the
literature, we have not found sufficient details to reproduce them
without going through the details given in this paper.  In
particular, we found it necessary to derive some apparently new results
for factorization for the form factor, which we give in App.\ \ref{app:ff}.

We collected together the results from different sources, and then
have results for the hard coefficient $H$, for the anomalous
dimensions $\gamma_j$ and $\gamma_K$, and for the CS-style evolution function
$\tilde{K}(\Tsc{b})$ at order $\alpha_s^3$.  The remaining perturbative
function is the small-$\Tsc{b}$ matching coefficient, which in all
cases is known to order $\alpha_s^2$.

There are several noteworthy observations to make here: On one hand,
approaches starting from calculations in collinear
  factorization and the form factor, which do not use
  explicit definitions of TMD functions, gave results for
all perturbative parts ($\gamma_j$, $\gamma_K$, $H$, and $C$-functions
in TMD factorization) without the
need to deal with TMD-specific issues such as how to regulate rapidity
divergences in the operator matrix elements defining TMD
  functions. This is a major advantage of such methods. Another
advantage is that the steps to obtain all perturbatively calculated
quantities are the same that are needed to calculate $\Tsc{q}{} \sim Q$
corrections (called the $Y$-terms; see also
Ref.~\cite{Stewart:2013faa} and references therein for other
approaches).  Thus, all relevant perturbative calculations are
included.  On the other hand, methods that specify clear TMD pdf
definitions also uniquely fix the definition of the hard part, $H$, up
to renormalization schemes. Without such definitions, there is
ambiguity in defining a hard part, as discussed in
Ref.~\cite{Catani:2000vq}.  However, as we have shown, the
ambiguity is completely resolved by appropriate manipulations applied
to results for the massless quark form factor despite there being no
explicit use of the definitions of the definitions of the TMD
functions.  Methods such as those of
\cite{Collins:2011qcdbook, GarciaEchevarria:2011rb, Gehrmann:2012ze,
  Gehrmann:2014yya, Grozin:2015kna, Echevarria:2016scs,
  Luebbert:2016itl, Li:2016ctv}, which begin with explicit TMD
definitions, have the advantages of allowing direct
calculations of the relevant quantities, and of allowing the
efficient realization even higher order calculations, as in
\citet{Li:2016ctv}, for some 
quantities, and they also leave open the possibility of studying
TMD correlation functions directly, even non-perturbatively.
A loss of a clear separation between hard parts and
correlation functions is a disadvantage of approaches rooted purely in
collinear factorization and large $\Tsc{q}{}$ methods.  Our hope is
that results from this article will allow the advantages of each
approach to be optimally exploited. In future work, this would include
in treatments of polarization-dependent effects, using spin dependent
matching coefficients such as those calculated recently in
Ref.~\cite{Gutierrez-Reyes:2017glx}.

Thirdly, we have extended the universality properties of the TMD functions by 
proving in App.~\ref{app:ff} that the anomalous dimensions labeled $\gamma_j$ 
(these are labeled $\gamma_F$ and $\gamma_D$ in, e.g., Ref.~\cite{Rogers:2015sqa})
are equal between TMD pdfs and TMD fragmentation functions to all orders.
In the past, fixed order calculations were suggestive of this result, but it can 
be now taken as a general theorem.

The compatibility that we have demonstrated between alternative formalisms, many of which appear very 
different on the surface, provides a highly non-trivial test of the general structure of TMD 
factorization. Also, at a practical level, this means that perturbative ingredients needed for implementing
TMD factorization are available at several loop order. This will be important for future efforts to implement TMD 
factorization phenomenologically in multiple and diverse contexts (for recent work, see~\cite{Bacchetta:2017gcc} and references therein).


\begin{acknowledgments}
  This work was supported in part by the
  U.S. Department of Energy under Grant No.\ DE-SC0013699. 
This work was also supported by the DOE Contract No.~DE-AC05-06OR23177,
under which Jefferson Science Associates, LLC operates Jefferson Lab.
This material is also based upon work supported by the U.S. Department
of Energy, Office of Science, Office of Nuclear Physics,
within the framework of the TMD Topical Collaboration.
  We thank C. Aidala and A. Idilbi for useful conversations. We thank 
  S. Catani and M. Grazzini for helpful discussions of their hard 
  parts and resummation scheme. We thank M. Diehl for very helpful comments on 
  the text.
\end{acknowledgments}

\appendix

\section{Results on quark form factor}
\label{app:ff}

In working with factorization for the quark form factor, some
complications arise concerning the phases of the various factors.  A
particular issue concerns the phases of the collinear factors and
their relation to the orientation (past- or future-pointing) of the
Wilson lines used in defining them. The phases give a possibility that
the anomalous dimensions have imaginary parts; their effects need to
be understood to give a correct relation between anomalous dimensions
and hard parts for the form factor and corresponding quantities for
the Drell-Yan and SIDIS cross sections.

This appendix gives the necessary results.  A primary tool is the
application of $TP$ invariance to relate amplitudes with past- and
future-pointing Wilson lines; this generalizes the method used
in Ref.\ \cite{Collins:2002kn} and Sec.\ 13.17.1 of
  \cite{Collins:2011qcdbook}
to relate parton densities between SIDIS and Drell-Yan.

\subsection{Definitions of on-shell wave functions}

In defining the form factor and the collinear factors, as used in
factorization properties like Eq.\ \eqref{eq:fact.sud.TL}, it is
necessary to extract the spin-dependence associated with the Dirac
wave functions of the external particles.  So here we define these
wave functions in terms of operator matrix elements.  The formulas are
standard, and are important in systematizing the application of $TP$
symmetry. 

Let $\ket{p,s}$ be the state of an incoming quark of momentum $p$ with
the spin part of the state defined by a label $s$.  We will leave
unstated the flavor of the quark for the moment. Its Dirac wave function is
defined to be
\begin{equation}
  \label{eq:wf.u}
  u\xleft( \ket{p,s} \right) = \frac{1}{\sqrt{Z}} \braket{ 0 | \psi(0) | p,s },
\end{equation}
where $Z$ is the residue of the on-shell pole of the quark's
propagator, and $\psi$ is the quark's field.  For an outgoing quark,
we use
\begin{equation}
  \label{eq:wf.ubar}
  \bar{u}\xleft( \bra{p,s} \right) = \frac{1}{\sqrt{Z}} \braket{ p,s | \bar{\psi}(0) | 0 },
\end{equation}
which can, of course, be derived from the hermitian conjugate of Eq.\
\eqref{eq:wf.u}.

For an antiquark, we indicate the state with an overbar,
$\ket{\overline{\strut p,s}}$, and define the wave function for an
incoming antiquark by
\begin{equation}
  \label{eq:wf.v}
  \bar{v}\xleft( \ket{\overline{\strut p,s}} \right) = 
  \frac{1}{\sqrt{Z}} \braket{ 0 | \bar{\psi}(0) | \overline{\strut p,s} },
\end{equation}
and for an outgoing antiquark by
\begin{equation}
  \label{eq:wf.vbar}
  v\xleft( \bra{\overline{\strut p,s}} \right) = 
  \frac{1}{\sqrt{Z}} \braket{ \overline{\strut p,s} | \psi(0) | 0 }.
\end{equation}

(Throughout we use the standard convention where an S-matrix element
is notated as $\braket{ \mbox{out} | \mbox{in} }$, with the out-state
as a bra and the in-state as a ket.)

\subsection{Definitions of scalar electromagnetic form factor}

For the time-like form factor for quark-antiquark annihilation,
$q_j(p_A) + \bar{q}_{\bar{\jmath}}(p_B) \to \gamma^*(q)$, the actual
amplitude is defined by
\begin{equation}
\label{eq:F.mu.is}
  \hat{F}^\mu_{\text{i.s.}}
  = \braket{ 0 | j^\mu(0) | p_A,s_A, \overline{\strut p_B,s_B}, \text{in} }.
\end{equation}
Here ``i.s.'' denotes ``initial-state''.
We choose coordinates such that the 3-momenta of the particles,
$\3{p}_A$ and $\3{p}_B$, are in the $+z$ and $-z$ directions. We
use light-front coordinates, defined for a vector $v$ by
$v=(v^+,v^-,\T{v})
= \left( (v^0+v^z)/\sqrt{2}, (v^0-v^z)/\sqrt{2}, \T{v} \right)$.

We define the scalar form factor $F_{\text{i.s.}}$ by
\begin{equation}
  \hat{F}^\mu_{\text{i.s.}}
  = \bar{v}_B \gamma^\mu u_A  ~ F_{\text{i.s.}}(Q^2) + \mbox{power-suppressed},
\end{equation}
where $\bar{v}_B$ and $u_A$ are the wave functions for the external
particles, and $Q^2=(p_A+p_B)^2$.
In the massless limit this is equivalent to 
\begin{equation}
   F_{\text{i.s.}}(Q^2)=
   -\frac{1}{4(1 - \epsilon) Q^2} {\rm Tr} \left( \slashed{p}_A \Gamma^\mu 
    \slashed{p}_B \gamma^\nu \right) \, ,
\end{equation}
where $\Gamma^\mu$ is the vertex function, i.e., the matrix element in
Eq.~\eqref{eq:F.mu.is} with the factors of $u_A$ and $\bar{v}_B$
omitted.

To relate the wave-function
structure to the approximation that naturally appears in
factorization, we define projection matrices
\begin{equation}
  P_A = \frac{1}{2} \gamma^- \gamma^+,
\qquad
  P_B = \frac{1}{2} \gamma^+ \gamma^-.
\end{equation}
Then to leading power in $Q$, the amplitude is
\begin{equation}
  \hat{F}^\mu_{\text{i.s.}}
  = \bar{v}_B P_A \gamma^\mu P_A u_A  ~F_{\text{i.s.}}(Q^2) 
    + \mbox{power-suppressed}.
\end{equation}

When the quark and antiquark are in the final state, so that the
process is $ \gamma^*(q) \to q_j(p_A) + \bar{q}_{\bar{\jmath}}(p_B)$, 
we have instead
\begin{equation}
  \hat{F}^\mu_{\text{f.s.}}
  = \braket{ p_A,s_A, \overline{\strut p_B,s_B}, \text{out} | j^\mu(0) | 0 },
\end{equation}
and
\begin{equation}
  \hat{F}^\mu_{\text{f.s.}}
  = \bar{u}_A P_B \gamma^\mu P_B v_B  ~ F_{\text{f.s.}}(Q^2) 
    + \mbox{power-suppressed}.
\end{equation}

For the space-like process, $\gamma^*(q) + q_j(p_A) \to q_j(p_B)$, with
an incoming and an outgoing quark, we have
\begin{align}
  \hat{F}^\mu_{\text{SL}}
  ={}& \braket{ p_B,s_B, \text{out} | j^\mu(0) | p_A,s_A, \mbox{in} }
\nonumber\\
  ={}& \bar{u}_B P_B \gamma^\mu P_A v_A  ~ F_{\text{SL}}(Q_E^2) 
    + \mbox{power-suppressed},
\end{align}
where $Q_E^2 = - (p_B-p_A)^2$. 

As is well known, the time-like form factors for incoming particles
and outgoing particles are equal, while the time-like form factor is
obtained by analytically continuing the space-like form factor to
$Q_E^2=-Q^2-i\epsilon$.
So we can write $F_{\text{i.s.}}(Q^2) = F_{\text{f.s.}}(Q^2) =
  F_{\text{SL}}(-Q^2-i\epsilon) = F(-Q^2-i\epsilon)$, where we no longer need
  labels to distinguish the different versions.

\subsection{Factorization and the definitions of collinear factors}

Factorization for the time-like form factor has the form
shown in
\eqref{eq:fact.sud.TL}, to which are to be added
  power-suppressed corrections if masses are non-zero.
Associated with it are statements of the 
dominant regions that contribute to the factors together with the
evolution equations \eqref{eq:evo.C}, \eqref{eq:C.rap.evol},
and \eqref{eq:KsudRG}.

To derive factorization in the case that the quark and antiquark are
incoming (such as in Drell-Yan scattering),
the Wilson lines in the definitions of the collinear factors
must be past pointing \cite[Ch.\ 10]{Collins:2011qcdbook}.  This is to
make it possible to deform the contour of integration over loop
momenta out of the Glauber region.  In this case the directions of the
Wilson lines match those of the corresponding quark or antiquark.

When, instead, the quark and antiquark are in the final state, the
Wilson lines are future pointing.  The collinear factors are therefore
potentially different than for the initial-state case.  We will later
use $TP$ invariance to show that they are in fact equal.

Finally, for the space-like case, it might appear natural to use a
mixture of past-pointing and future-pointing Wilson lines, to
correspond to the physical situation of having one incoming quark and
one outgoing quark.  But in fact they can be chosen to be all
future-pointing.  The reasoning is the same as for factorization in
SIDIS (\cite{Collins:1997sr}, \cite[Sec.\
12.14.3]{Collins:2011qcdbook}).  The Wilson lines could also be chosen
to be all past-pointing.  The choice for them to be future-pointing
enables the results for the space-like form factor to match the
results for corresponding graphs in SIDIS.

We define a Wilson line in direction $n$ as the operator
\begin{equation}
  W_n = P \exp\xleft[ -ig_0\int_0^\infty n\cdot A^{(0)}(\lambda n) \diff{\lambda} \right],
\end{equation}
where $P$ denotes path-ordering, $g_0$ is the bare coupling and
$A^{(0)}$ is the bare gluon field, a matrix on color space.  We now define
the collinear factor $C_{j,\text{ i.s., past}}$ for an initial-state
quark of flavor $j$ with past-pointing Wilson lines.  In the method
explained in \cite[Ch.\ 10]{Collins:2011qcdbook}, we need auxiliary
soft factors
\begin{equation}
  S_{\text{i.s.}}(y_1,y_2) = \braket{ 0 | W_{n_2}W_{n_1}^\dagger | 0 } ,
\end{equation}
which in fact only depend on the rapidity difference $y_1-y_2$.
Here $n_1$ and $n_2$ denote the following directions, of rapidities
$y_1$ and $y_2$, in light-front coordinates:
\begin{equation}
  n_1 = - (1, -e^{-2 y_1},\T{0}), 
\qquad
  n_2 = - ( - e^{2 y_2}, 1, \T{0}).
\end{equation}
We will be working with limits $y_1\to+\infty$ and $y_2\to-\infty$,
when $n_1$ and $n_2$ become past-pointing directions corresponding to
the incoming quark and antiquark.

The collinear factors have an extra auxiliary direction in their
definition; it is given a rapidity $y$.  We define the collinear
factor, $C_{j,\text{ i.s., past}}(\zeta_A,\mu)$ as used in the
factorization theorem, by
\begin{widetext}
\begin{equation}
\label{eq:Cj.def}
  \lim_{y_1\to\infty, y_2\to-\infty} P_A
     \braket{ 0 | W_{n_2}\psi_{j,(0)}(0) | p_A,s_A }
     \sqrt{ \frac{ S_{\text{i.s.}}(y_1,y) }
             { S_{\text{i.s.}}(y_1,y_2) ~ S_{\text{i.s.}}(y,y_2) }
     }
     \times Z_j
  = C_{j,\text{ i.s., past}}(\zeta_A,\mu) ~ P_A u(\ket{p_A,s_A}) .
\end{equation}
Given that QCD is invariant under rotations and parity inversion, the
spin dependence is only as given by the last factor on the right,
leaving the scalar collinear factor $C_j$.  The quantity $Z_j$ is a UV
renormalization factor, which in fact equals the quantity $\exp\xleft(
\frac12 D -i \frac{\pi}{2} E + \frac12 \ln\frac{Q^2}{\mu^2} E
\right)$, with $D$ and $E$ as used in Sec.\ \ref{sec:ff}.  The
collinear factor depends on the rapidity of the auxiliary direction
$y$ via the parameter
\begin{equation}
  \zeta_A = 2 (p_A^+)^2e^{-2y}.
\end{equation}
This, and the corresponding $\zeta_B$ for the antiquark's collinear
factor, may be set equal to $|Q^2|$.

Almost the same definition gives the collinear factor
$C_{\bar{\jmath},\text{ i.s., past}}(\zeta_B,\mu)$ for the antiquark.
The directions must then be adjusted to be compatible with the chosen
direction for the antiquark's momentum, which also exchanges the roles
of the directions $n_1$ and $n_2$.  This gives
\begin{equation}
  \lim_{y_1\to\infty, y_2\to-\infty} 
     \braket{ 0 | \bar{\psi}_{j,(0)}(0) W_{n_1}^{\dagger} |
       \overline{\strut p_B,s_B} }
     P_A
     \sqrt{ \frac{ S_{\text{i.s.}}(y,y_2) }
             { S_{\text{i.s.}}(y_1,y_2) ~ S_{\text{i.s.}}(y_1,y) }
     }
     \times Z_j
  =
  C_{\bar{\jmath},\text{ i.s., past}}(\zeta_B,\mu)
    ~ \bar{v}(\ket{\overline{\strut p_B,s_B}}) ~ P_A .
\end{equation}
\end{widetext}
By charge-conjugation invariance the antiquark and quark collinear
factors are equal:
\begin{equation}
  C_{\bar{\jmath},\text{ i.s., past}}(\zeta,\mu)
  = C_{j,\text{ i.s., past}}(\zeta,\mu).
\end{equation}

As already indicated, there are three other versions of the
collinear factors that need to be considered in turn: 
We may replace the past-pointing Wilson lines by
future-pointing Wilson lines, and, independently, we may change the
initial-state quark to a final-state quark.  For example, in the case
of a final-state quark, the quark matrix element (times projector) is
replaced by
\begin{equation}
\label{eq:Cj.f.s.OME}
  \braket{ p_A,s_A | \bar{\psi}_{j,(0)}(0) W_{n_2}^\dagger | 0 } P_B,
\end{equation}
and the wave function factor by
\begin{equation}
  \bar{u}(\bra{p_A,s_A}) P_B.
\end{equation}

We will next find the relations between the four versions of the
collinear factor.  (Some relations are elementary consequences of
applying hermitian conjugation, of course.)

\subsection{Using $TP$ symmetry etc to relate collinear factors for
  different cases}
\label{sec:TP}

Since a $TP$ transformation reverses both space and time coordinates,
we will use $TP$ invariance to relate collinear factors with
initial-state and final-state Wilson lines.  Since both $T$ and $P$
separately reverse the 3-momentum of a state, the combined $TP$
operations preserves momentum.

We let $U_{TP}$ be the anti-unitary operator for $TP$ transformations
on state space.  To specify its action on the fields, we choose to use
the Dirac representation of his matrices:
\begin{equation}
  \gamma^0 =
  \begin{pmatrix}
    1 & 0 \\ 0 & -1
  \end{pmatrix},
\qquad
  \gamma^i =
  \begin{pmatrix}
    0 & \sigma^i \\ -\sigma^i & 0.
  \end{pmatrix}.
\end{equation}
Then the $TP$ transformation of the quark and antiquark fields is
given by
\begin{align}
\label{eq:psi.TP}
  U_{TP} \psi(x) U_{TP}^{-1} ={}& \gamma_{TP} \psi(-x),
\\
  U_{TP} \bar{\psi}(x) U_{TP}^{-1} ={}& \bar{\psi}(-x) \gamma_{TP} ,
\end{align}
where
\begin{equation}
  \gamma_{TP} = i \gamma^1\gamma^3\gamma^0, 
\end{equation}
which is hermitian, imaginary and antisymmetric, and is its own
inverse. 

The inverse transformations acquire a minus sign, which will be
important to our calculations:
\begin{align}
  U_{TP}^{-1} \psi(x) U_{TP} ={}& - \gamma_{TP} \psi(-x),
\\
  U_{TP}^{-1} \bar{\psi}(x) U_{TP} ={}& - \bar{\psi}(-x) \gamma_{TP} ,
\end{align}

It can be shown, from the effect of $TP$ on the gluon fields, that a
$TP$ transformation simply reverses the direction of a Wilson line:
\begin{equation}
  U_{TP} W_n U_{TP}^{-1} = W_{-n}.
\end{equation}

Although a $TP$ transformation preserves the momentum of a quark (or
other) state, it changes the spin in a way governed by the field
transformations. For example,
\begin{equation}
\label{eq:TP.u}
  u\xleft( U_{TP}\ket{p,s} \right)
  =
  \gamma_{TP} u\xleft( \ket{p,s} \right)^*.
\end{equation}
This simply follows from Eqs.\ \eqref{eq:wf.u} and \eqref{eq:psi.TP},
as do similar equations for the other varieties of wave function.

We now use a $TP$ transformation to relate collinear factors with
past-pointing and future-pointing Wilson lines.  We start from the
definition \eqref{eq:Cj.def}, but with the quark state $\ket{p,s}$
replaced by $U_{TP}\ket{p,s}$.  The following chain of argument
relates the quark matrix element to the complex conjugate of a matrix
element with a reversed Wilson line. We drop some subscripts from the
notation, since they do not matter here.
\begin{align}
  & P_A \braket{ 0 | W_n\psi(0) U_{TP} | p,s }
\nonumber\\
  &= 
  P_A \braket{ 0 | U_{TP}U_{TP}^{-1} W_n U_{TP}U_{TP}^{-1} \psi(0)  U_{TP} | p,s }
\nonumber\\
  &=
  P_A \braket{ 0 | U_{TP} W_{-n} (-\gamma_{TP}) \psi(0) | p,s }
\nonumber\\
  &=
  \gamma_{TP} \biggl[ P_A \braket{ 0 | W_{-n} \psi(0) | p,s } \biggr]^* .
\end{align}
The antilinearity of $U_{TP}$ gives the complex conjugation in the
last line.  It also gives a sign reversal of the imaginary matrix
$\gamma_{TP}$, when this numerical matrix is taken from a position on
the right of $U_{TP}$ to the left.  A similar argument shows that the
soft factors in \eqref{eq:Cj.def} with their past-pointing Wilson
lines equal the complex conjugate of the soft factors with
future-pointing Wilson lines.  Hence as regards the left-hand-side of
\eqref{eq:Cj.def}, we have
\begin{widetext}
\begin{equation}
\label{eq:TP.Cj}
  \mbox{L.h.s.\ of \eqref{eq:Cj.def} with past W.L.\ and state $U_{TP}\ket{p,s}$}
  = \gamma_{TP}
    \times \biggl( 
          \mbox{L.h.s.\ with future W.L.\ and state $\ket{p,s}$}
           \biggr)^*
\end{equation}
For the right-hand-side of \eqref{eq:Cj.def}, with the state
$U_{TP}\ket{p,s}$ we have
\begin{equation}
  C_{j,\text{ i.s., past}}(\zeta,\mu) ~ P_A ~ u(U_{TP}\ket{p,s})
  =
  C_{j,\text{ i.s., past}}(\zeta,\mu) ~ \gamma_{TP} ~ P_A \biggl[ u(\ket{p,s}) \biggr]^*,
\end{equation}
\end{widetext}
from \eqref{eq:TP.u}. But the right-hand side of \eqref{eq:TP.Cj} equals
\begin{equation}
  \gamma_{TP} \biggl[ 
      C_{j,\text{ i.s., future}}(\zeta,\mu) P_A  u(\ket{p,s}) 
  \biggr]^*.
\end{equation}
We deduce that 
\begin{equation}
  C_{j,\text{ i.s., future}}(\zeta,\mu)
  =
  \biggl[ C_{j,\text{ i.s., past}}(\zeta,\mu) \biggr]^* .
\end{equation}
That is, changing Wilson lines between future- and past-pointing
causes a complex conjugation of the collinear factor.

We also need to relate this to the collinear factor for a final-state
quark. Now the quark matrix element \eqref{eq:Cj.f.s.OME} used for a
final-state quark is the hermitian conjugate of the one for an
initial-state quark.  But hermitian-conjugation leaves the location of
the Wilson line unchanged.  From this and a little further algebra we
deduce that 
\begin{equation}
  C_{j,\text{ f.s., future}}(\zeta,\mu)
  =
  \biggl[ C_{j,\text{ i.s., future}}(\zeta,\mu) \biggr]^*,
\end{equation}
and hence
\begin{equation}
  C_{j,\text{ f.s., future}}(\zeta,\mu)
  =
  C_{j,\text{ i.s., past}}(\zeta,\mu).
\end{equation}
Thus we have equal collinear factors if the directions of the Wilson
lines match the quarks, and a complex conjugate when they are
opposite.

For a reference collinear factor, notated simply $C_j$, we use
\begin{equation}
  C_j = C_{j,\text{ i.s., past}} = C_{j,\text{ f.s., future}}, 
\end{equation}
and then the other two cases are
\begin{equation}
  C_{j,\text{ i.s., future}} = C_{j,\text{ f.s., past}} = C_j^*.
\end{equation}

\subsection{Phases for anomalous dimensions, etc}
\label{sec:phases}

From the results so far (and the established factorization
properties), we have evolution equations of the form \eqref{eq:evo.C}
and \eqref{eq:C.rap.evol}.  We know, from explicit calculations, that
a collinear factor can and does have a non-trivial phase. So the
anomalous dimension functions $\gamma_j$ and $\gamma_K$ might also
have phases.

To show that they are in fact real, we start from the observation that
from the above results, the space-like form factor obeys
\begin{equation}
  F^{\rm SL} = H^{\rm SL} |C_j|^2,
\end{equation}
with the absolute value squared of the collinear factor. The
space-like form factor $F^{\rm SL}$ is real, and therefore so is the
corresponding hard factor. Going to the massless case, we replace
$|C_j|^2$ by its counterterm, and, just as we had in
  \eqref{eq:ln.H} for the time-like case, we have
\begin{equation}
  \ln F^{\rm SL}
  =
  \ln H^{SL} + D^{\rm SL} + \ln\frac{Q_E^2}{\mu^2} E^{\rm SL}
\quad
\text{(massless),}
\end{equation}
where we have used a superscript ``SL'' on $D$ and $E$ because we have
not yet established their identity with those used with the time-like
form factor.  Since $F$ is real, we find its pole terms, captured in
the $D$ and $E$ terms, are also real.

Now we analytically continue $F$ to the time like case.  The pole part
analytically continues to 
\begin{equation}
  D^{\rm SL} + \ln\frac{-Q^2-i\epsilon}{\mu^2} E^{\rm SL}
=
  D^{\rm SL} -i\pi E^{\rm SL} + \ln\frac{Q^2}{\mu^2} E^{\rm SL},
\end{equation}
and the finite part $H$ continues to its value for the time-like case:
\begin{equation}
  \label{eq:HSud.TL.SL}
  H^{\text{Sud, TL}}(Q^2) = H^{\text{Sud, SL}}(-Q^2-i\epsilon).
\end{equation}

Comparison of the above equations with \eqref{eq:ln.H} for the
time-like case, shows that $D^{\rm SL}$ and $E^{\rm SL}$
are equal to the original $D$ and $E$, and that these functions are
real. It also follows that $\gamma_j$ and $\gamma_K$ are real, since
they can be computed from simple derivatives of $D$ and $E$.

\section{Correspondence with methods of \citet{Li:2016ctv, Li:2016axz}}
\label{app:corr.Li.et.al}

\citet{Li:2016ctv} have made an important calculation at order $a_s^3$
of the kernel of the rapidity RG equation of their soft factor.  As we
will show in this section, their kernel in fact exactly equals the
$\tilde{K}$ function of CSS2.  Their definition appears to be quite
different to that of $\tilde{K}$, so the equality is far from obvious,
which leads to the proof given in this section.

Furthermore their TMD factorization formula includes an explicit soft
factor, similarly to case for the TMD factorization formula in CSS1
prior to CSS1's process-dependent redefinitions.  In this section, we
will also show, following Refs.\ \cite{GarciaEchevarria:2011rb,
  Collins:2012uy}, how to convert the factorization formula and TMD
parton densities used by \citet{Li:2016ctv} to the CSS2 form, which in turn are the same 
as those of \citet{GarciaEchevarria:2011rb} (see Ref.~\cite{Collins:2012uy}), thereby
giving a standardized set of parton densities common to most recent
formalisms.

The version of TMD factorization that is used in Ref.\
\cite{Li:2016ctv} uses a regulator of rapidity divergences defined by
\citet{Li:2016axz}.  The hard factor agrees with Eqs.~(\ref{eq:HDY},\ref{eq:HSIDIS}), since it
corresponds to virtual graphs for the on-shell quark form factor with
collinear and soft subtractions.  After allowing for differences in
conventions for an overall normalization factor, we find that their
factorization formula differs from the CSS2 version
\eqref{eq:fact.CSS2} simply by the replacement of the factors
$\tilde{f}_{j/A} \tilde{f}_{\bar{\jmath}/B}$ by
\begin{widetext}
\begin{align}
\label{eq:Li.fact}
           \lim_{\nu\to\infty}   B(x_A,b_T; \mu, \nu/(x_AP_A^+),a_s(\mu))
              ~ B(x_B,b_T; \mu, \nu/(x_BP_B^-),a_s(\mu))
              ~ S(b_T; \mu, \nu, a_s(\mu)).
\end{align}
Here $\nu$ is the rapidity regulator parameter, the $B$ factors are beam
functions with zero bin subtractions\footnote{''Zero bin subtractions'' in SCET refers to the removal of overlap between the beam function 
and soft gluons.} applied, and $S$ is a soft
factor.  In the operator definitions, $\nu$ is implemented as follows:
In $S$, the vertices on the left and the right of the final-state cut
have their relative positions changed from the standard value
$b=(0,0,\T{b})$, as used in CSS2, to $b=(ib_0/\nu,ib_0/\nu,\T{b})$, in
$(+,-,T)$ coordinates. This is called an exponential rapidity regulator. In the beam functions, only the component of
position separation that is zero in the unregulated quantity is
replaced in this fashion --- in Ref.\ \cite{Li:2016axz} see Eq.\ (33),
as compared with the unregulated form (13), although we did not find 
explicit definitions of the regulated beam factors.  Because the shift is
applied equally to $+$ and $-$ coordinates there is an implicit choice
of Lorentz frame, similar to the choice of the rapidity for the
non-lightlike vector in CSS2. The dependence of each beam function on
$\nu$ and $x_AP_A^+$ or $x_BP_B^-$ is by $\nu/(x_AP_A^+)$ or
$\nu/(x_BP_B^-)$ only, and this is determined by Lorentz covariance and
the specific implementation of the regulator.

This exponential rapidity regulator \cite{Li:2016ctv,Li:2016axz} does not have any effect on purely virtual graphs. In a
full QCD treatment including infrared physics, the virtual graphs need separate regulators, but in the combination
used in \eqref{eq:Li.fact}, this extra regulator may be removed, since
the associated divergences cancel in the product.  Using such a
regulator would be important in determining the correspondence with
matrix elements that can be calculated in non-perturbative models
(including the use of lattice gauge theory).  However the calculations
we are interested in are all in a purely massless theory with on-shell
external partonic targets.  In that case, the would-be-divergent
integrals for the virtual graphs for the product of collinear and soft
factors are scale-free and hence are consistently zero.  This is
exactly the same as for the graphs for the bare collinear factors
for the massless quark form factor 
that we examined in Sec.\ \ref{sec:form.factor.factorization}; the
graphs are the same.  There remain \MSbar{} renormalization factors
that can be determined from the other graphs.
Hence, as regards calculations, only
graphs with real emission are considered, which is what is done in the
calculations to order $a_s^3$ in Ref.\ \cite{Li:2016ctv}. 

A possible definition of TMD parton densities (e.g.,
\cite{Li:2016axz}) would be as the beam functions in
\eqref{eq:Li.fact}, with the regulator preserved, but apparently with
the asymptotic behavior as $\nu\to\infty$ extracted --- see Eq.\ (2) of Ref.\
\cite{Li:2016ctv}.  However, as already mentioned, the soft factor
does not have a phenomenologically independent appearance.  So
following the method of \citet{GarciaEchevarria:2011rb}, we can
redefine the TMD parton densities by absorbing a factor of $\sqrt{S}$
into each, and then removing the regulator:
\begin{equation}
  \label{eq:pdf.Li.Zhu}
  \tilde{f}_{j/A}(x_A,\Tsc{b};\zeta_A,\mu)
  = \lim_{\nu\to\infty}   B(x_A,b_T; \mu, \nu/x_AP_A^+,a_s(\mu))
              ~ \sqrt{ S(b_T; \mu, \nu, a_s(\mu)) } ,
\end{equation}
\end{widetext}
and similarly for $\tilde{f}_{\bar{\jmath}/B}$.  Here $\zeta_A=2(x_AP_A^+)^2$
with a corresponding definition $\zeta_B=2(x_BP_B^-)^2$ for the other
parton density (in, for example, Drell-Yan scattering). The choice of frame for defining the non-boost
invariant quantities $P_A^+$ and $P_B^-$ is determined by the
implementation of the regulator $\nu$.  As in CSS2, we have
$\zeta_A\zeta_B=Q^4$, and without loss of generality we can set $\zeta_A=\zeta_B=Q^2$
after applying evolution equations.

The above confirms that the only differences between the TMD pdfs of \cite{Li:2016ctv,Li:2016axz} and 
\cite{Collins:2012uy} are in the details for implementing rapidity cutoffs.
Thus, the resulting TMD parton densities are the same as those of
CSS2 \cite{Collins:2012uy}. These TMD parton densities are therefore
universal between the different formalisms, and should be considered as
standard.  In particular, they are independent of the exact method by
which rapidity divergences are regulated and canceled, at least as
regards the different approaches in Refs.\ \cite{Li:2016axz,
  GarciaEchevarria:2011rb, Echevarria:2016scs, Collins:2011qcdbook,
  Chiu:2011qc, Chiu:2012ir, Luebbert:2016itl}.  In all cases, in the
limit that the regulator(s) are removed, there is a
collinear factor that is a matrix element of the standard
gauge-invariant operator for TMD parton densities, with exactly
light-like Wilson lines.  This is multiplied by a combination of a UV
renormalization factor, soft factors, and possibly a factor
implementing zero-bin subtractions.

Now, for determining the CSS2 $\tilde{K}$, the relevant evolution
equation in \citet{Li:2016ctv} is the rapidity RG equation
\cite{Chiu:2011qc, Chiu:2012ir}, for dependence on $\nu$.  We wish to
relate this to $\tilde{K}$, which is defined as a derivative of a
parton density with respect to a different variable $\zeta$.  We first
observe that the dependence of the beam function on $\nu$ is by the
ratio $\nu/(x_AP_A^+)$, and so the $\nu$-dependence is determined by the
dependence on $P_A^+$ and hence on $\zeta_A$. Now to get a finite limit as
$\nu\to\infty$ in \eqref{eq:pdf.Li.Zhu} we must have
\begin{multline}
\label{eq:RRG.B}
  \frac{\partial}{\partial\nu}\ln B(x_A,b_T; \mu, \nu/x_AP_A^+,a_s(\mu))
\\
 = - \frac{\partial}{\partial\nu} \frac{1}{2} \ln S(b_T; \mu, \nu, a_s(\mu))].
\end{multline}
Hence
\begin{equation}
\label{eq:K.gammaR}
  \tilde{K} = \frac{\partial\ln \tilde{f}}{\partial \sqrt{\zeta}}
            = - \lim_{\nu\to\infty} \frac{\partial\ln B}{\partial\nu}
            = \frac{1}{2} \lim_{\nu\to0} \frac{\partial\ln \tilde{S}}{\partial\nu}
            = \gamma_R,
\end{equation}
which quantity was given in (15) of \citet{Li:2016axz}, and was used
in (4) of \citet{Li:2016ctv}.  Equation (23) of Ref.\
\cite{Li:2016axz} is essentially a version of our \eqref{eq:RRG.B}, as
is Eq.\ (2.16) of Ref.\ \cite{Chiu:2012ir}.

When we set $\mu$ to its standard value $b_0/\Tsc{b}{}$ we get equality of
$\tilde{K}$ with the quantity $\gamma_r$ used in Ref.\ \cite{Li:2016ctv}:
\begin{equation}
  \left. \tilde{K}(b_T,\mu,a_s(\mu)) \right|_{\mu\mapsto b_0/b_T}
  = \gamma_r(a_s).
\end{equation}
The value of $\gamma_r$ is given numerically to order $a_s^3$ in (9) of
\citet{Li:2016ctv}, and hence also gives CSS2's $\tilde{K}$.


\bibliography{jcc}

\end{document}